\documentclass[10pt]{article}

\usepackage{amsmath}
\usepackage{amssymb}

\usepackage{graphicx}

\usepackage{cite}

\usepackage{color} 

\usepackage[T1]{fontenc}

\usepackage[labelfont=bf,labelsep=period,justification=raggedright]{caption}

\makeatletter
\renewcommand{\@biblabel}[1]{\quad#1.}
\makeatother

\newtheorem{mydef}{Definition}

\date{}

\begin{document}

\begin{flushleft}
{\Large
\textbf{Information geometry, simulation and complexity in Gaussian random fields}
}
\\
Alexandre L. M. Levada$^{1}$
\\
\bf{1} Computing Department, Federal University of S\~ao Carlos, S\~ao Carlos, SP, Brazil
$\ast$ E-mail: Corresponding alexandre@dc.ufscar.br
\end{flushleft}


\section*{Abstract}
Random fields are useful mathematical objects in the characterization of non-deterministic complex systems. A fundamental issue in the evolution of dynamical systems is how intrinsic properties of such structures change in time. In this paper, we propose to quantify how changes in the spatial dependence structure affect the Riemannian metric tensor that equips the model's parametric space. Defining Fisher curves, we measure the variations in each component of the metric tensor when visiting different entropic states of the system. Simulations show that the geometric deformations induced by the metric tensor in case of a decrease in the inverse temperature are not reversible for an increase of the same amount, provided there is significant variation in the system's entropy: the process of taking a system from a lower entropy state A to a higher entropy state B and then bringing it back to A, induces a natural intrinsic one-way direction of evolution. In this context, Fisher curves resemble mathematical models of hysteresis in which the natural orientation is pointed by an arrow of time.


\section{Introduction}

Over the years, the study and characterization of complex systems have become a major research topic in many areas of science~\cite{CPLX,Sibani}. Part of this massive interest is due to a common requirement in the modeling and analysis of several natural phenomena existing in the world around us: to understand how relationships between pieces of information give rise to collective behaviors among different scale levels of a system \cite{Strogatz,Barabasi}. Reasons for the appearance of this complexity are countless and are not completely known. Often, in complex systems, the interaction between the components is highly non-linear and/or non-deterministic, which brings several challenges that prevent us from getting a better understanding of the underlying processes that govern the global behavior of such structures \cite{Newman2003,Boccaletti}.

With the growing volume of data that is being produced in the world these days, the notion of information is more present and relevant in any scale of modern society \cite{BigData}. In this scenario, where data plays a central role in science, an essential step in order to learn, understand and assess the rules governing complex phenomena that are part of our world is not only the mining of relevant symbols along this vast ocean of data, but especially the identification and further classification of these patterns \cite{Hastie,Mining}. After the pieces of information are put together and the relationship between them is somehow uncovered, a clearer picture start to emerge, as in the solution of an intricate puzzle. In this paradigm, computational tools for data analysis and simulations are a fundamental component of this data-driven knowledge discovery process \cite{escience}. 

In this context, random fields are particularly interesting mathematical structures \cite{RandomFields}. First, it is possible to replace the usual statistical independence assumption by a more realistic conditional independence hypothesis \cite{Besag1974}. In other words, unlike most classical stochastic models, we can incorporate the dependence between random variables in a formal and elegant way. This is a key aspect when one needs to study how local interactions can lead to the emergence of global effects. Second, if we constrain the size of the maximum clique to be two, that is, we assume only binary relationships, then we have a pairwise interaction Markov model, which is mathematically tractable \cite{Besag1974,MLaPP}. Finally, considering that the coupling parameter is invariant and isotropic, all the information regarding the spatial dependence structure of the random field is conveyed by a single scalar parameter, from now on denoted by $\beta$. In the physics literature, this parameter is referred as the inverse temperature of the system, and plays an important role in statistical mechanics and thermodynamics \cite{Boltzmann,Gibbs,Ising1925,Heisenberg,Spitzer,PottsWu,Albeverio}. 

Random fields have been used with success in several areas of science from a long time ago \cite{Montroll1941,Onsager,KnotWu,Liu,Merks,Geman1984,Sabelfeld}. Recently, information geometry has emerged as an unified approach in the study and characterization of the parametric spaces of random variables by combining knowledge from two distinct mathematical fields: differential geometry and information theory \cite{Amari1985,Amari2000,Frieden2004,Dodson}. However, most information geometry studies are focused in the classical assumption of independent samples drawn from exponential family of distributions \cite{Nielsen,Pistone,Amari2001}. Little is known about information geometry on random fields, more precisely, about how the geometric properties of the parametric space of these models are characterized. Although some related work can be found in the literature \cite{Amari1992,Janke,Xavier,Zanardi}, there are still plenty of room for contributions in this field. 

Along centuries many researchers have studied the concept of time \cite{Campisi,TimeArrow,Haddad}. During our investigations, some questions that motivated this research were based on the relation between time and complexity: what are the causes to the emergence of complexity in dynamical systems? Is it possible to measure complex behavior along time? What is time? Why does time seem to flow in one single direction? How to characterize time in a complex system? We certainly do not have definitive answers to all these questions, but in an attempt to study the effect of time in the emergence of complexity in dynamical systems, this paper proposes to investigate an information-theoretic approach to understand these phenomena in random fields composed by Gaussian variables. Our study focuses on the information theory perspective, motivated by the connection between Fisher information and the geometric structure of stochastic models, provided by information geometry.

The main goal of this paper is to characterize the information geometry of Gaussian random fields, through the derivation of the full metric tensor of the model's parametric space. Basically, we want to sense each component of this Riemannian metric $g$ as we perform positive and negative displacements in the inverse temperature ``axis'' in order to measure the geometric deformations induced to the underlying manifold (parametric space). It is known that when the inverse temperature parameter is zero, the model degenerates to a regular Gaussian distribution, whose parametric space exhibit constant negative curvature (hyperbolic geometry) \cite{Sueli}. It is quite intuitive to think that the shape and topology of the parametric space has a deep connection with the distances between random fields operating in different regimes, which is crucial in characterizing the behavior of such systems. To do so, we propose to investigate how the metric tensor components change while the system navigates through different entropic states.

In summary, we want to track all the deformations in the metric tensor from an initial configuration A, in which temperature is infinite ($\beta = 0$), to a final state B, in which temperature is much lower. Additionally, we want to repeat this process of measuring the deformations induced by the metric tensor, but now starting at B and finishing at A. If the sequence of deformations A$\rightarrow$B is different from the sequence of deformations B$\rightarrow$A, it means that the process of taking the random field from an initial lower entropic state A to a final higher entropic state B and bring it back to A induces a natural intrinsic one way direction of evolution: an arrow of time. In practical terms, our proposal consists in using information geometry as a mathematical tool to measure the emergence of an intrinsic notion of time in random fields in which temperature is allowed to deviate from infinity \cite{Levada2014}.  

Since we are restraining our analysis only to Gaussian random fields, which are mathematically tractable, exact expressions for the components of the full metric tensor are explicitly derived. Computational simulations using Markov-Chain Monte Carlo algorithms \cite{Hastings1970,Swendsen1987,Wolff1989,gilks1993,smith1993,Roberts1996,Landau2000,Chib2004} validate our hypothesis that the emergence of an arrow of time in random fields is possibly a consequence of asymmetric deformations in the metric tensor of the statistical manifold when the inverse temperature parameter is disturbed. However, in searching for a solution to this main problem in question, two major drawbacks have to be overcome: 1) the information equality does not hold for $\beta > 0$, which means that we have two different versions of Fisher information; and 2) the computation of the expected Fisher information (the components of the metric tensor) requires knowledge of the inverse temperature parameter for each configuration of the random field. The solution for the first sub-problem consists in deriving not one but two possible metric tensors: one using type-I Fisher information and another using type-II Fisher information. For the second sub-problem our solution was to perform maximum pseudo-likelihood estimation in order to accelerate computation by avoiding calculations with the partition function in the joint Gibbs distribution. Besides, these two sub-problems share an important connection: it has been verified that the two types of Fisher information play a fundamental role in quantifying the uncertainty in the maximum pseudo-likelihood estimation of the inverse temperature parameter $\beta$ through the definition of the asymptotic variance of this estimator \cite{Levada2014}.

In the following, we describe a brief outline of the paper. In Section 2 we define the pairwise Gaussian-Markov random field (GMRF) model and discuss some basic statistical properties. In addition, we provide an alternative description of the evolving complex system (random field) as a non-deterministic finite automata in which each cell may assume an infinite number of states. In Section 3 the complete characterization of the metric tensor of the underlying Riemannian manifold in terms of Fisher information is detailed. Section 4 discusses maximum pseudo-likelihood, a technique for estimating the inverse temperature parameter given a single snapshot of the random field. Section 5 presents the concept of Fisher curve, a geometrical tool to study the evolution of complex systems modelled by random fields by quantifying the deformations induced to the parametric space by the metric tensor. Section 6 shows the computational simulations using Markov Chain Monte Carlo (MCMC) algorithms, the obtained results and some final remarks. Finally, Section 7 presents the conclusions of the paper.

\section{The Random Field Model}

The objective of this section is to introduce the random field model, characterizing some basic statistical properties. Gaussian random fields are important models in dealing with spatially dependent continuous random variables, once they provide a general framework for studying non-linear interactions between elements of a stochastic complex system along time. One of the main advantages of these models is the mathematical tractability, which allows us to derive exact closed-form expressions for two relevant quantities in this investigation: 1) estimators for the inverse temperature parameter; and 2) the expected Fisher information matrix (the Riemannian metric of the underlying parametric space manifold). According to the Hammersley-Clifford theorem \cite{Hammersley1971}, which states the equivalence between Gibbs random fields (global models) and Markov random fields (local models) it is possible to characterize an isotropic pairwise Gaussian random field by a set of local conditional density functions (LCDF's), avoiding computations with the joint Gibbs distribution (due to the partition function).

\begin{mydef}
An isotropic pairwise Gaussian Markov random field regarding a local neighborhood system $\eta_{i}$ defined on a lattice $S=\left\{ s_{1}, s_{2}, \ldots , s_{n} \right\}$ is completely characterized by a set of $n$ local conditional density functions $p( x_{i} | \eta_{i}, \vec{\theta} )$, given by:
\end{mydef}

\begin{equation}
	p\left( x_{i} | \eta_{i}, \vec{\theta} \right) = \frac{1}{\sqrt{2\pi\sigma^2}}exp\left\{-\frac{1}{2\sigma^{2}} \left[ x_{i} - \mu - \beta \sum_{j \in \eta_{i}} \left( x_{j} - \mu \right) \right]^{2} \right\}
	\label{eq:GMRF}
\end{equation} with $\vec{\theta} = (\mu, \sigma^{2}, \beta)$ the parameters vector, where $\mu$ and $\sigma^{2}$ are respectively the expected value (mean) and the variance of the random variables in the field, and $\beta$ is the inverse temperature or coupling parameter, which is responsible for controlling the global spatial dependence structure of the system. Note that if $\beta = 0$, the model degenerates to the usual Gaussian model for independent random variables.


\begin{mydef}
A model $p\left(\mathbf{X}| \vec{\theta} \right)$ belongs to the $K$ parametric exponential family if it can be expressed as:
\end{mydef}

\begin{equation}
	p\left(\mathbf{X}| \vec{\theta} \right) = exp\left\{\sum_{j=1}^{K}c_{j}( \vec{\theta} )T_{j}\left( \mathbf{X} \right) + d( \vec{\theta} ) + S\left( \mathbf{X} \right) \right\}
\end{equation} where $\vec{c} = \left(c_{1}(\vec{\theta}), c_{2}(\vec{\theta}), \ldots, c_{k}(\vec{\theta}) \right)$ is a vector of natural parameters, $\vec{T} = \left( T_{1}\left(\mathbf{X} \right), T_{2}\left(\mathbf{X} \right), \ldots, T_{k}\left(\mathbf{X} \right) \right)$ is vector of natural sufficient statistics, $d(\vec{\theta})$ is an arbitrary function of the parameters and $S\left(\mathbf{X} \right)$ is an arbitrary function of the observations. A model is called curved if the  dimensionality $K$ of both $\vec{c}$ and $\vec{T}$ (number of natural sufficient statistics) is greater than the dimensionality $D$ of the parameter vector $\vec{\theta}$ (number of parameters in the model). For instance, considering a sample $\mathbf{X} = \{x_{1}, x_{2}, \ldots, x_{n}\}$ of the isotropic pairwise Gaussian Markov random field model in which $\Delta$ denotes the support of the neighborhood system (i.e, 4, 8, 12, etc.), we can express the joint conditional distribution, which is the basis for the definition of the pseudo-likelihood function \cite{Besag1974}, as:

\begin{align}
	& F\left(\mathbf{X}| \eta_{i}, \vec{\theta} \right) = \left( 2\pi\sigma^2\right)^{-n/2}exp\left\{ -\frac{1}{2\sigma^2}\sum_{i=1}^{n}\left[ \left(x_{i} - \mu \right) - \beta \sum_{j\in\eta_i}\left( x_{j} - \mu \right)  \right]^2  \right\} \\ \nonumber 
	& = \left( 2\pi\sigma^2\right)^{-n/2}exp\left\{ -\frac{1}{2\sigma^2}\sum_{i=1}^{n}\left[ x_{i}^{2} -2x_{i}\mu + \mu^2 - 2\beta\sum_{j\in\eta_i}(x_{i} - \mu)(x_{j} - \mu)  \right. \right. \\ \nonumber & \hspace{6cm} + \left. \left. \beta^2 \sum_{j\in\eta_i}\sum_{k\in\eta_i}(x_{j} - \mu)(x_{k} - \mu)\right] \right\} \\ \nonumber
	& = exp\left\{-\frac{n}{2}log(2\pi\sigma^2) - \frac{1}{2\sigma^2}\sum_{i=1}^{n}x_{i}^2 + \frac{\mu}{\sigma^2}\sum_{i=1}^{n} x_{i} - \frac{n\mu^2}{2\sigma^2} \right. \\ \nonumber & \hspace{4cm} + \left. \frac{\beta}{\sigma^2} \left[ \sum_{i=1}^{n}\sum_{j\in\eta_i} x_{i}x_{j} - \mu \Delta \sum_{i=1}^{n} x_{i} -\mu\sum_{i=1}^{n}\sum_{j\in\eta_i} x_{j} + \Delta \mu^2 n \right] \right\} \\ \nonumber
	&\times  exp\left\{ -\frac{\beta^2}{2\sigma^2}\left[ \sum_{i=1}^{n} \sum_{j\in\eta_i}\sum_{k\in\eta_i}x_{j}x_{k} - \mu \Delta \sum_{i=1}^{n} \sum_{j\in\eta_i}x_{j} - \mu \Delta \sum_{i=1}^{n}\sum_{k\in\eta_i}x_{k} + \Delta^2 \mu^2 n \right] \right\} \\ \nonumber \\ \nonumber
	& = exp\left\{ -\frac{n}{2}\left[ log(2\pi\sigma^2) + \frac{\mu^2}{\sigma^2} \right] + \frac{\beta\Delta\mu^2 n}{\sigma^2}\left[ 1 - \frac{\beta\Delta}{2} \right] \right\} \\ \nonumber 
	&\times exp\left\{ \left[ \frac{\mu}{\sigma^2}\left(1 - \beta\Delta\right) \right]\sum_{i=1}^{n}x_{i} -\frac{1}{2\sigma^2}\sum_{i=1}^{n}x_{i}^2 + \frac{\beta}{\sigma^2}\sum_{i=1}^{n}\sum_{j\in\eta_i}x_{i}x_{j} \right. \\ \nonumber & \left. \hspace{4cm} - \left[ \frac{\beta\mu}{\sigma^2}(1 - \beta\Delta)\right]\sum_{i=1}^{n}\sum_{j\in\eta_i}x_{j} - \frac{\beta}{2\sigma^2}\sum_{i=1}^{n}\sum_{j\in\eta_i}\sum_{k\in\eta_i}x_{j}x_{k}  \right\}
\end{align}

By observing the above equation, it is possible to identify the following correspondence: 

\begin{align}
	\vec{c} = \left( \left[ \frac{\mu}{\sigma^2}\left(1 - \beta\Delta\right) \right], -\frac{1}{2\sigma^2}, \frac{\beta}{\sigma^2}, -\left[ \frac{\beta\mu}{\sigma^2}(1 - \beta\Delta)\right], - \frac{\beta}{2\sigma^2} \right) \\ \nonumber
	\vec{T} = \left( \sum_{i=1}^{n}x_{i}, \sum_{i=1}^{n}x_{i}^2, \sum_{i=1}^{n}\sum_{j\in\eta_i}x_{i}x_{j}, \sum_{i=1}^{n}\sum_{j\in\eta_i}x_{j}, \sum_{i=1}^{n}\sum_{j\in\eta_i}\sum_{k\in\eta_i}x_{j}x_{k}  \right) 
\end{align} with $S(\mathbf{X}) = 0$ and 

\begin{equation}
	d(\vec{\theta}) = -\frac{n}{2}\left[ log(2\pi\sigma^2) + \frac{\mu^2}{\sigma^2} \right] + \frac{\beta\Delta\mu^2 n}{\sigma^2}\left[ 1 - \frac{\beta\Delta}{2} \right]
\end{equation}

Note that the model is a member of the curved exponential family, since even though the parametric space is a 3D manifold, the dimensionality of $\vec{c}$ and $\vec{T}$ is more than that (there is a total of 5 different natural sufficient statistics, more than one for each parameter). Once again, notice that for $\beta = 0$, the mathematical structure is reduced to the traditional Gaussian model where both vectors $\vec{c}$ and $\vec{T}$ are 2 dimensional, perfectly matching the dimension of the parameters vector $\vec{\theta} = \left(\mu, \sigma^2 \right)$:

\begin{equation}
	F\left(\mathbf{X}| \vec{\theta} \right) = exp\left\{ \frac{\mu}{\sigma^2}\sum_{i=1}^{n}x_{i} -\frac{1}{2\sigma^2}\sum_{i=1}^{n}x_{i}^2  -\frac{n}{2}\left[ log(2\pi\sigma^2) + \frac{\mu^2}{\sigma^2} \right] \right\}
\end{equation} where now we have $S(\mathbf{X}) = 0$ and:

\begin{align}
	\vec{c} = \left( \frac{\mu}{\sigma^2}, -\frac{1}{2\sigma^2} \right) \\ \nonumber
	\vec{T} = \left( \sum_{i=1}^{n}x_{i}, \sum_{i=1}^{n}x_{i}^2 \right) \\ \nonumber
	d(\vec{\theta}) = -\frac{n}{2}\left[ log(2\pi\sigma^2) + \frac{\mu^2}{\sigma^2} \right]
\end{align}

Hence, from a geometric perspective, as the inverse temperature parameter in a random field deviates from zero, a complex deformation process transforms the underlying parametric space (a 2D manifold) into a completely different structure (a 3D manifold). It has been shown that the geometric structure of regular exponential family distributions exhibit constant curvature. It is also known that from an information geometry perspective \cite{Amari2000,Kass1989}, the natural Riemannian metric of these probability distribution manifolds is given by the Fisher information matrix. However, little is known about information geometry on more general statistical models, such as random field models. In this paper, our primary objective is to study, from an information theory perspective, how changes in the inverse temperature parameter affect the metric tensor of the Gaussian Markov random field model. The idea is that by measuring these components (Fisher information) we are capturing and quantifying an important complex deformation process induced by the metric tensor into the parametric space as temperature is disturbed. Our main goal is to investigate how displacements in the inverse temperature parameter direction (``$\beta$ axis'') affect the metric tensor and as a consequence, the geometry of the parametric space of random fields.

\subsection{Random Fields Dynamics as Non-deterministic Cellular Automata}

The evolution of a random field from a given initial configuration is a dynamical process that can be viewed as the simulation of a non-deterministic cellular automata in which each cell has a probability to accept a new behavior depending on the behaviors of the neighboring cells in the grid. Essentially, this is what is done by Markov Chain Monte Carlo algorithms to perform random walks throughout the state space of a random field model during a sampling process. 

In this paper a cellular automata is considered as a continuous dynamical system defined on a discrete space (2D rectangular lattice). The system is governed by local rules defined in terms of the neighborhood of the cells in a way that these laws describe how the cellular automata evolves in time.

\begin{mydef}
A discrete-space cellular automata can be represented as a sextuple $\Omega = \left( S, I, f, f_{0}, \eta, \phi  \right)$, where \cite{Odemir}:
\end{mydef}  

\begin{itemize}
	\item $S$ is a n-dimensional lattice of the Euclidean space $\Re^n$, consisting of cells $s_{i}$, $i \in N$;
	\item $I$ is a set of states for each cell (in our model $I = \Re$ is an infinite continuous set that represents the outcome of a Gaussian random variable to express an infinite number of possible behaviors);
	\item An output function $f:S \times N \rightarrow I$ maps the state of a cell $s_{i}$ at a discrete time $t$, denoted by $f(s_{i},t)$;
	\item $f_{0}$ is an initial configuration (in our model it is a random configuration generated by the outputs of $|S|$ independent gaussian variables);
	\item A neighborhood function $\eta: S \rightarrow S^{\Delta}$ yields every cell $s_{i}$ to a finite sequence $\eta_{i} \in S^{\Delta}$ so that $\eta_{i} = ( s_{i_j} )_{j=1}^{\Delta}$ has $\Delta$ distinct cells $s_{j}$ ($\Delta$ is the support of the neighborhood system);
	\item A transition function $\phi:S^{\Delta} \rightarrow S$ describes the rules governing the dynamics of every cell $s_{i} \in S$ so that:
	\begin{equation}
		f(s_{i}, t+1) = \phi\left( \left( f(s_{j},t) \right)_{j\in\eta_{i}} \right) = \phi\left( \delta_{i} \right)
	\end{equation}		
\end{itemize}

Thus, the resulting cellular automata characterization for our particular random field model is given by: $S$ is the 2D rectangular lattice, $I$ is the real line (to allow each cell to express an infinite number of possible behaviors), an output $f$ is performed by sampling from the probability density function of a given cell $s_{i}$ (the LCDF of the random field model as given by equation \ref{eq:GMRF}), the neighborhood function $\eta$ is the usual Moore neighborhood (the 8 nearest neighbors) and the transition function $\phi$ is defined in terms of the Metropolis-Hastings acceptance rate. To do so, let $P$ be defined as:

\begin{equation}
	P = \frac{p \left( \tilde{x_i}|\eta_i, \vec{\theta} \right)}{p \left( x_i | \eta_i, \vec{\theta} \right)}
\end{equation} where both $\tilde{x_i}$ and $x_i$ are two different outputs for a cell $s_i$. In other words, $\tilde{x_i}$ and $x_i$ denote two possible values for $f(s_i, t)$. Let $P' = min\{1, P \}$ be the minimum value between 1 and $P$. Then, the transition function is given by:

\begin{equation}
	\phi \left( \begin{array}{ccc}
	x_1 & x_2 & x_3 \\ 
	x_4 & x_i & x_5 \\ 
	x_6 & x_7 & x_8
	\end{array} \right) = \begin{cases}
							\tilde{x_i} & \text{with prob. } P', \\
							x_i & \text{with prob. } 1-P'
						 \end{cases}
\end{equation} where the parameter $P$ used to compute $P'$ can be written as:

\begin{align}
	\label{eq:P}
	P & = exp\Bigg\{ -\frac{1}{2\sigma^2}\Bigg[ (\tilde{x_i} - \mu)^2 - (x_i - \mu)^2 \\ \nonumber & \hspace{3cm} + 2\beta \Bigg( \sum_{j\in\eta_i}(\tilde{x_i} - \mu)(x_j - \mu) - \sum_{j\in\eta_i}(x_i - \mu)(x_j - \mu) \Bigg) \Bigg]  \Bigg\}
\end{align}

\begin{figure}[h]
	\center
	\includegraphics[scale=0.4]{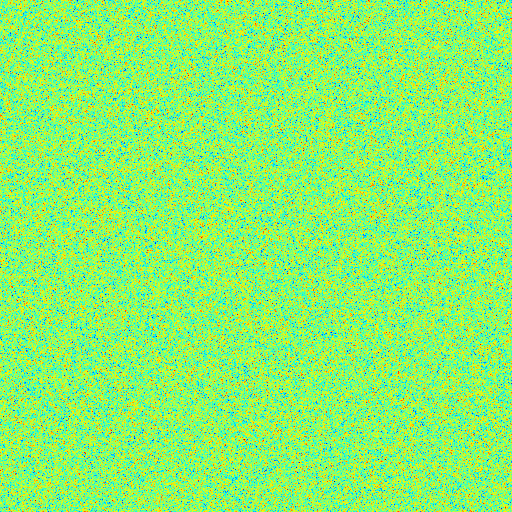} 
	\includegraphics[scale=0.4]{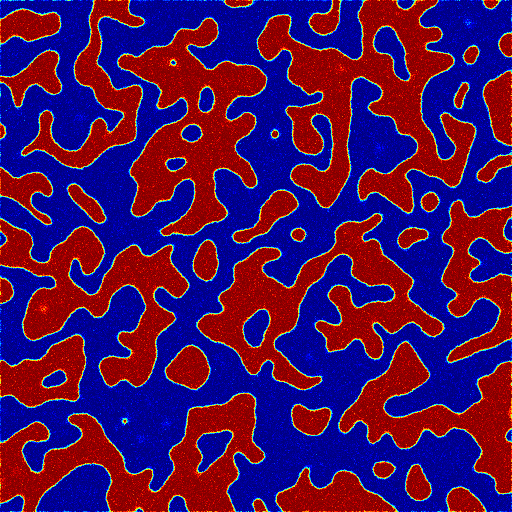} 	
	\caption{{\bf Different configurations representing a non-deterministic cellular automata.}
	Global versus local adjustment according to the spatial dependence structure induced by the inverse temperature parameter $\beta$.}
	\label{fig:automata}
\end{figure}

Some observations are important at this point. First, the rule for the non-deterministic automata can be put in words as: generate a new candidate for the behavior of the cell $s_i$, compute $P$ and accept the new behavior with probability $P'$ or keep the old behavior with probability $1 - P'$. The crucial part however is the analysis of the transition function in terms of the spatial dependence structure of the random field, controlled by the inverse temperature parameter. Note that, when $\beta \rightarrow 0$, the second term of equation \eqref{eq:P} (inside the parenthesis) vanishes, indicating that the transition function favors behaviors that are similar to the global one, indicated in this model by the expected value or simply the $\mu$ parameter. In this scenario, new behaviors are considered probable if they fit the global one. On the other hand, when $\beta$ grows significantly, this second term, which is a measure of local  adjustment, becomes increasingly relevant to the transition function. In these situations, the cells try to adjust their behaviors to the behavior of the nearest neighbors, ignoring the global structure. Figure \ref{fig:automata} illustrates two distinct configurations regarding the evolution of a Gaussian random field. The left one corresponds to the initial random configuration in which the inverse temperature parameter $\beta$ is zero. The right image is the configuration obtained after 200 steps of evolution for $\beta$ starting at zero and with regular and fixed increments of $\Delta\beta = 0.005$ in each iteration. Different colors encode different behaviors for the cells in the grid. Note the major difference between the two scenarios described above.

In summary, our main research goal with this paper is to investigate how changes in the inverse temperature parameter affect the transition function of a non-deterministic cellular automata modeled according to a Gaussian random field. This investigation is focused in the analysis of Fisher information, a measure deeply related to the geometry of the underlying random field model's parametric space, since it provides the basic mathematical tool for the definition of the metric tensor (natural Riemannian metric) of this complex statistical manifold.

\section{The Metric Tensor for Gaussian Random Fields}

In this section, we discuss how information geometry can be applied in the characterization of the statistical manifold of Gaussian random fields by the definition of the proper Riemannian metric, given by the Fisher information matrix. Information geometry has been a relevant research area since the pioneering works of Shun'ichi Amari \cite{Amari1985,Amari2000} in the 80's, developed by the application of theoretical differential geometry methods to the study of mathematical statistics. Since then, this field has been expanded and successfully explored by researchers in a wide range of science areas, from statistical physics and quantum mechanics to game theory and machine learning. 

Essentially, information geometry can be viewed as a branch of information theory that provides a robust and geometrical treatment to most parametric models in mathematical statistics (belonging to the exponential family of distributions). Within this context, it is possible to investigate how two distinct independent random variables from the same parametric model are related in terms of intrinsic geometric features. For instance, in this framework it is possible to measure distances between two Gaussian random variables $X \sim N(\mu_x,\sigma_{x}^{2})$ and $Y \sim N(\mu_y, \sigma_{y}^{2})$.

Basically, when we analyse isolated random variables (that is, they are independent), the scenario is extensively known, with the underlying statistical manifolds being completely characterized. However, little is known about the scenario in which we have several variables interacting with each other (in other words, the inverse temperature parameter is not null). In geometric terms, this imply the emergence of an extra dimension in the statistical manifold, and therefore, in the metric tensor. We will see in the following subsections that the emergence of this inverse temperature parameter ($\beta$) strongly affects all the components of the original metric tensor. 

Suppose $p(x|\vec{\theta})$ is a statistical model belonging to the exponential family, where $\vec{\theta}$ denotes the parameters vector of the model. Then, the collection of all admissible vectors $\vec{\theta}$ defines the parametric space $\Theta$, which has shown to be a Riemannian manifold. Moreover, it has been shown that in the Gaussian case, the underlying manifold is a surface with constant negative curvature, defining its geometry as hyperbolic \cite{Kass1989,Dodson}. Since the parametric space $\Theta$ is not an Euclidean space, it follows that the manifold is curved. Thus, to make the computation of distances and arc lengths in $\Theta$ possible, it is necessary to express an infinitesimal displacement $ds$ in the manifold in an adaptive or locally way. Roughly speaking, that is the reason why a manifold must be equipped with a metric tensor, which is the mathematical structure responsible for the definition of inner products in the local tangent spaces. With the metric tensor it is possible to express the square of an infinitesimal displacement in the manifold, $ds^2$, as a function of an infinitesimal displacement in the tangent space, which in case of a 2D manifold is given by a vector $[du, dv]$. Assuming a matrix notation we have:

\begin{equation}
	ds^2 = \begin{bmatrix} du & dv \end{bmatrix} \begin{bmatrix} A & B \\ B & C \end{bmatrix}\begin{bmatrix} du \\ dv \end{bmatrix} = A du^2 + 2Bdudv + C dv^2
\end{equation} where the matrix of coefficients $A$, $B$, e $C$ is the metric tensor. If the metric tensor is a positive definite matrix, the manifold is is known as Riemannian. Note that in the Euclidean case, where the metric tensor is the identity matrix (since the space is flat), we have the known Pitagorean relation $ds^2 = du^2 + dv^2$.

\subsection{Fisher information}

Since its definition, in the works of Sir Ronald Fisher \cite{Fisher}, the concept of Fisher information has been present in an ubiquitous manner throughout mathematical statistics, playing an important role in several applications, from numerical estimation methods based on the Newton-Raphson iteration to the definition of lower bounds in unbiased estimation (Cramer-Rao lower bound). More recently, with the development of information geometry, another fundamental role of Fisher information in statistical models has been discovered: it defines intrinsic geometric properties of the parametric space of a model, by characterizing the metric tensor of the respective manifold. In other words, the Fisher information matrix is the natural Riemannian metric of the manifold (parametric space), given a statistical model.

Roughly speaking, Fisher information can be thought as a likelihood analog to entropy, which is often used as a measure of uncertainty, but it is based in probability, not likelihood. Basically, in the context of information theory, Fisher information measures the amount of information a random sample conveys about an unknown parameter. 

\begin{mydef}
	Let $p(X;\vec{\theta})$ be a probability density function where $\vec{\theta} = (\theta_1, \ldots, \theta_n) \in \Theta$ is the parametric vector. The Fisher information matrix, which is the natural Riemannian metric of the parametric space, is defined as:
\end{mydef}

\begin{equation}
	\left\{ I(\vec{\theta}) \right\}_{ij} = E\left[ \left(\frac{\partial}{\partial\theta_i} log~p(X; \vec{\theta}) \right)\left(\frac{\partial}{\partial\theta_j} log~p(X;\vec{\theta}) \right) \right], \text{~~~~ for } i,j=1,\ldots,n
\end{equation} 

It is known from the statistical inference theory that information equality holds for independent observations from the regular exponential family of distributions. In other words, it is possible to compute the expected Fisher information matrix of a model by two different but equivalent ways (since the integration and differentiation operators can be interchangeable), defining the condition known as the information equality: 

\begin{equation}
	E\left[ \left(\frac{\partial}{\partial\theta_i} log~p(X; \vec{\theta}) \right)\left(\frac{\partial}{\partial\theta_j} log~p(X; \vec{\theta}) \right) \right] = -E\left[ \frac{\partial^2}{\partial\theta_i \partial\theta_j} log~p(X; \vec{\theta}) \right]
\end{equation}

In this investigation we replace $p(X; \vec{\theta})$ by the local conditional density function of an isotropic pairwise Gaussian random field (equation \ref{eq:GMRF}). More details on how this LCDF is used to build the pseudo-likelihood function are presented in the next sections of the paper.

However, what we observe is that, given the intrinsic spatial dependence structure of random field models, induced by the existence of an inverse temperature parameter, information equality is not a natural condition. In general, when the inverse temperature parameter gradually drifts apart from zero (temperature deviates from infinity), this notion of information "equilibrium" fails. Thus, in dealing with random field models, we have to consider two different versions of Fisher information, from now on denoted by type-I (due to the first derivative operator in the log likelihood function) and type-II (due to the second derivative operator). Eventually, when certain conditions are satisfied, these two values of information converge to a unique bound. One trivial condition for the information equality is to have $\beta=0$, which means an infinite temperature (there is no induced spatial dependence structure since the variables are independent and the model degenerates to a regular exponential family density).

Therefore, in random fields, these two versions of Fisher information play distinct roles, especially in quantifying the uncertainty in the estimation of the inverse temperature parameter, as we will see in future sections.

\subsection{The Riemannian Metric: Characterizing the Metric Tensor}

In this section we present the derivation of all components of the metric tensor $g$ in an isotropic pairwise Gaussian Markov random field model. The complete characterization of both versions of the metric tensor, using type-I and type-II Fisher information is discussed in details. For purposes of notation, we define these tensors as:

\begin{equation}
	g^{(1)}(\vec{\theta}) = \left( \begin{array}{ccc}
	I_{\mu\mu}^{(1)}(\vec{\theta}) & I_{\mu\sigma^2}^{(1)}(\vec{\theta}) & I_{\mu\beta}^{(1)}(\vec{\theta}) \\ 
	I_{\sigma^2\mu}^{(1)}(\vec{\theta}) & I_{\sigma^2\sigma^2}^{(1)}(\vec{\theta}) & I_{\sigma^2\beta}^{(1)}(\vec{\theta}) \\ 
	I_{\beta\mu}^{(1)}(\vec{\theta}) & I_{\beta\sigma^2}^{(1)}(\vec{\theta}) & I_{\beta\beta}^{(1)}(\vec{\theta})
	\end{array} \right)
\end{equation} and

\begin{equation}
	g^{(2)}(\vec{\theta}) = \left( \begin{array}{ccc}
	I_{\mu\mu}^{(2)}(\vec{\theta}) & I_{\mu\sigma^2}^{(2)}(\vec{\theta}) & I_{\mu\beta}^{(2)}(\vec{\theta}) \\ 
	I_{\sigma^2\mu}^{(2)}(\vec{\theta}) & I_{\sigma^2\sigma^2}^{(2)}(\vec{\theta}) & I_{\sigma^2\beta}^{(2)}(\vec{\theta}) \\ 
	I_{\beta\mu}^{(2)}(\vec{\theta}) & I_{\beta\sigma^2}^{(2)}(\vec{\theta}) & I_{\beta\beta}^{(2)}(\vec{\theta})
	\end{array} \right)
\end{equation} where $g^{(1)}(\vec{\theta})$ is the type-I Fisher information matrix and $g^{(2)}(\vec{\theta})$ is the type-II Fisher information matrix.

\subsubsection{The $g^{(1)}(\vec{\theta})$ Metric Tensor}

In the following, we proceed with the definition of the type-I Fisher information matrix. The first component of $g^{(1)}(\vec{\theta})$ is given by:

\begin{equation}
	I_{\mu\mu}^{(1)}(\vec{\theta}) = E\left[ \left(\frac{\partial}{\partial\mu} log~p(X; \vec{\theta}) \right)\left(\frac{\partial}{\partial\mu} log~p(X; \vec{\theta}) \right) \right]
\end{equation} where $p(X; \vec{\theta})$ is the replaced by the LCDF of the Gaussian random field, given by equation \eqref{eq:GMRF}. Plugging the equations and computing the derivatives leads to:

\begin{align}
	I_{\mu\mu}^{(1)}(\vec{\theta}) & = E\left\{ \frac{1}{\sigma^2}\left(1 - \beta\Delta \right)^2 \frac{1}{\sigma^2}\left[ \left(x_i - \mu \right) - \beta \sum_{j\in\eta_i}\left(x_j - \mu \right) \right]^2 \right\} \label{eq:mu_mu_1} \\ \nonumber & = \frac{1}{\sigma^2}\left(1 - \beta\Delta \right)^2 E\left\{ \frac{1}{\sigma^2} \left[ \left(x_i - \mu\right)^2 - 2\beta\sum_{j\in\eta_i}\left( x_i - \mu \right)\left( x_j - \mu \right) \right. \right. \\ \nonumber & \hspace{4cm} \left. \left. + \beta^2 \sum_{j\in\eta_i}\sum_{k\in\eta_i}\left( x_j - \mu \right)\left(x_k - \mu  \right) \right] \right\} \\ \nonumber & = \frac{\left(1 - \beta\Delta \right)^2}{\sigma^2} \left[ 1 - \frac{1}{\sigma^2}\left(  2\beta\sum_{j\in\eta_i}\sigma_{ij} - \beta^2\sum_{j\in\eta_i}\sum_{k\in\eta_i}\sigma_{jk} \right) \right]
\end{align} where $\Delta$ denotes the support of the neighborhood system (in our case $\Delta = 8$ since we have a second-order system), $\sigma_{ij}$ denotes the covariance between the central variable $x_i$ and one of its neighbors $x_j \in \eta_i$ and $\sigma_{jk}$ denotes the covariance between two variables $x_j$ and $x_k$ in the neighborhood $\eta_i$. The second component of the $g^{(1)}(\vec{\theta})$ metric tensor is:

\begin{equation}
	I_{\mu\sigma^2}^{(1)}(\vec{\theta}) = E\left[ \left(\frac{\partial}{\partial\mu} log~p(X; \vec{\theta}) \right)\left(\frac{\partial}{\partial\sigma^2} log~p(X; \vec{\theta}) \right) \right]
\end{equation} which leads to:

\begin{align}
	\label{eq:g12}
	I_{\mu\sigma^2}^{(1)}(\vec{\theta}) & = \frac{(1 - \beta\Delta)}{2\sigma^6}E\left\{\left[ \left( x_i - \mu \right) - \beta\sum_{j\in\eta_i}\left(x_j - \mu \right) \right]^3 \right\} \\ \nonumber & \hspace{3cm} -\frac{(1 - \beta\Delta)}{2\sigma^4}E\left\{ \left( x_i - \mu \right) - \beta\sum_{j\in\eta_i}\left(x_j - \mu \right) \right\}	
\end{align}

Note that second term of equation \eqref{eq:g12} is zero since:

\begin{equation}
	E\left[ x_i - \mu \right] - \beta\sum_{j\in\eta_i}E\left[ x_j - \mu\right] = 0 - 0 = 0
\end{equation} 

The expansion of the first term in \eqref{eq:g12} leads to:

\begin{align}
	\label{eq:g12_exp}
	E\left\{\left[ \left( x_i - \mu \right) - \beta\sum_{j\in\eta_i}\left(x_j - \mu \right) \right]^3 \right\} & = E\left[ \left( x_i - \mu \right)^3 \right] \\ \nonumber &- 3\beta\sum_{j\in\eta_i}E\left[ (x_i - \mu) (x_i - \mu) (x_j - \mu) \right] \\ \nonumber & + 3\beta^2 \sum_{j\in\eta_i}\sum_{k\in\eta_i}E\left[ (x_i - \mu) (x_j - \mu) (x_k - \mu) \right] \\ \nonumber & - \beta^3 \sum_{j\in\eta_i}\sum_{k\in\eta_i}\sum_{l\in\eta_i}E\left( (x_j - \mu) (x_k - \mu) (x_l - \mu) \right]
\end{align}

Note that the first term of \eqref{eq:g12_exp} is zero for Gaussian random variables since every central moment of odd order is null. According to the Isserlis' theorem \cite{isserlis1918}, it is trivial to see that in fact all the other terms are null. Therefore, $I_{\mu\sigma^2}^{(1)}(\vec{\theta})=0$. We now proceed to the third component of $g^{(1)}(\vec{\theta})$, defined by:

\begin{equation}
	I_{\mu\beta}^{(1)}(\vec{\theta}) = E\left[ \left(\frac{\partial}{\partial\mu} log~p(X; \vec{\theta}) \right)\left(\frac{\partial}{\partial\beta} log~p(X; \vec{\theta}) \right) \right]
\end{equation}

Replacing the equations and manipulating the resulting expressions leads to:

\begin{align}
	I_{\mu\beta}^{(1)}(\vec{\theta}) & = \frac{(1 - \beta\Delta)}{\sigma^4}\Bigg\{ E\left[ (x_i - \mu) (x_i - \mu) (x_j - \mu) \right]  \\ \nonumber & - 2\beta\sum_{j\in\eta_i}\sum_{k\in\eta_i}E\left[ (x_i - \mu) (x_j - \mu) (x_k - \mu) \right] \\ \nonumber & + \beta^2 \sum_{j\in\eta_i}\sum_{k\in\eta_i}\sum_{l\in\eta_i}E\left[ (x_j - \mu) (x_k - \mu) (x_l - \mu) \right] \Bigg\}
\end{align}

Once again, all the higher-order moments are a product of an odd number of Gaussian random variables so by the Isserlis's theorem they all vanish, resulting in $I_{\mu\beta}^{(1)}(\vec{\theta}) = 0$. For the next component, $I_{\sigma^2\mu}^{(1)}(\vec{\theta})$, we have:

\begin{equation}
	I_{\sigma^2\mu}^{(1)}(\vec{\theta}) = E\left[ \left(\frac{\partial}{\partial\sigma^2} log~p(X; \vec{\theta}) \right)\left(\frac{\partial}{\partial\mu} log~p(X; \vec{\theta}) \right) \right] = 0
\end{equation} since $I_{\mu\sigma^2}^{(1)}(\vec{\theta})=0$ and changing the order of the product does not affect the expected value. Proceeding to the fifth component of the metric tensor $g^{(1)}(\vec{\theta})$ we have to compute:

\begin{equation}
	I_{\sigma^2 \sigma^2}^{(1)}(\vec{\theta}) = E\left[ \left(\frac{\partial}{\partial\sigma^2} log~p(X; \vec{\theta}) \right)\left(\frac{\partial}{\partial\sigma^2} log~p(X; \vec{\theta}) \right) \right]
\end{equation} which is given by:

\begin{align}
	I_{\sigma^2 \sigma^2}^{(1)}(\vec{\theta}) & = E\left\{ \left[ -\frac{1}{2\sigma^2} + \frac{1}{2\sigma^4}\left( x_i - \mu - \beta\sum_{j\in\eta_i}(x_j - \mu) \right) \right]^2 \right\} \\ \nonumber & = \frac{1}{4\sigma^4} - \frac{1}{2\sigma^6}E\left\{ \left[ (x_i - \mu) - \beta\sum_{j\in\eta_i}(x_j - \mu) \right]^2 \right\} \\ \nonumber & \hspace{1cm} + \frac{1}{4\sigma^8}E\left\{ \left[ (x_i - \mu) - \beta\sum_{j\in\eta_i}(x_j - \mu) \right]^4 \right\}
\end{align}

In order to simplify the calculations, we expand each one of the expected values separately. The first expectation leads to the following equality:

\begin{equation}
	E\left\{ \left[ (x_i - \mu) - \beta\sum_{j\in\eta_i}(x_j - \mu) \right]^2 \right\} =  \sigma^2 - 2\beta\sum_{j\in\eta_i}\sigma_{ij} + \beta^2 \sum_{j\in\eta_i}\sum_{k\in\eta_i}\sigma_{jk} 
\end{equation}

In the expansion of the second expectation term note that:

\begin{align}
	& E\left\{ \left[ (x_i - \mu) - \beta\sum_{j\in\eta_i}(x_j - \mu) \right]^4 \right\} = E\left[ (x_i - \mu)^4 \right] - 4\beta\sum_{j\in\eta_i}E\left[ (x_i - \mu)^3 (x_j - \mu) \right] \nonumber \\ & \hspace{3cm} + 6\beta^2 \sum_{j\in\eta_i}\sum_{k\in\eta_i}E\left[ (x_i - \mu)^2 (x_j - \mu) (x_k - \mu) \right] \\ \nonumber & \hspace{3cm} - 4\beta^3 \sum_{j\in\eta_i}\sum_{k\in\eta_i}\sum_{l\in\eta_i} E\left[(x_i - \mu) (x_j - \mu) (x_k - \mu) (x_l - \mu) \right] \\ \nonumber & \hspace{3cm} + \beta^4 \sum_{j\in\eta_i}\sum_{k\in\eta_i}\sum_{l\in\eta_i}\sum_{m\in\eta_i}E\left[(x_j - \mu) (x_k - \mu) (x_l - \mu) (x_m - \mu) \right]
\end{align} leading to five novel expectation terms. Using the Isserlis' theorem for gaussian distributed random variables, it is possible to express the higher-order moments as functions of second-order moments. Therefore, after some algebra we have:

\begin{align}
	\label{eq:sigma_sigma_1}
	I_{\sigma^2 \sigma^2}^{(1)}(\vec{\theta}) = \frac{1}{2\sigma^4} & - \frac{1}{\sigma^6}\left[ 2\beta\sum_{j\in\eta_i}\sigma_{ij} - \beta^2 \sum_{j\in\eta_i}\sum_{k\in\eta_i}\sigma_{jk} \right] + \frac{1}{\sigma^8}\left[ 3\beta^2 \sum_{j\in\eta_i}\sum_{k\in\eta_i}\sigma_{ij}\sigma_{ik} \right. \\ \nonumber \\ \nonumber & \hspace{2cm} \left. - \beta^3 \sum_{j\in\eta_i}\sum_{k\in\eta_i}\sum_{l\in\eta_i}\left( \sigma_{ij}\sigma_{kl} + \sigma_{ik}\sigma_{jl} + \sigma_{il}\sigma_{jk} \right) \right. \\ \nonumber & \hspace{2cm} \left. + \beta^4 \sum_{j\in\eta_i}\sum_{k\in\eta_i}\sum_{l\in\eta_i}\sum_{m\in\eta_i}\left( \sigma_{jk} \sigma_{lm} + \sigma_{jl}\sigma_{km} + \sigma_{jm}\sigma_{kl} \right)  \right] 
\end{align}

The next component of the metric tensor is:

\begin{equation}
	I_{\sigma^2 \beta}^{(1)}(\vec{\theta}) = E\left[ \left(\frac{\partial}{\partial\sigma^2} log~p(X; \vec{\theta}) \right)\left(\frac{\partial}{\partial\beta} log~p(X; \vec{\theta}) \right) \right]
\end{equation} which is given by:

\begin{align}
	 I_{\sigma^2 \beta}^{(1)}(\vec{\theta}) = & E\left\{ \left[ -\frac{1}{2\sigma^2} + \frac{1}{2\sigma^4}\left( (x_i - \mu) - \beta\sum_{j\in\eta_i}(x_j - \mu)  \right)^2 \right] \times \right. \\ \nonumber & \hspace{3cm} \left. \left[ \frac{1}{\sigma^2}\left((x_i - \mu) - \beta\sum_{j\in\eta_i}(x_j - \mu)  \right)\left( \sum_{j\in\eta_i}(x_j - \mu) \right) \right] \right\} \\ \nonumber & = -\frac{1}{2\sigma^4} E\left\{  \left[ (x_i - \mu) - \beta\sum_{j\in\eta_i}(x_j - \mu) \right]\left[ \sum_{j\in\eta_i}(x_j - \mu) \right] \right\} \\ \nonumber & \hspace{2cm} + \frac{1}{2\sigma^6}E\left\{\left[ (x_i - \mu) - \beta\sum_{j\in\eta_i}(x_j - \mu) \right]^3 \left[ \sum_{j\in\eta_i}(x_j - \mu) \right]  \right\}
\end{align}


The first expectation can be simplified to:

\begin{equation}
	E\left\{  \left[ (x_i - \mu) - \beta\sum_{j\in\eta_i}(x_j - \mu) \right]\left[ \sum_{j\in\eta_i}(x_j - \mu) \right] \right\} = \sum_{j\in\eta_i}\sigma_{ij} - \beta\sum_{j\in\eta_i}\sum_{k\in\eta_i}\sigma_{jk}
\end{equation}

The expansion of the second expectation leads to:

\begin{align}
	& E\left\{\left[ (x_i - \mu) - \beta\sum_{j\in\eta_i}(x_j - \mu) \right]^3 \left[ \sum_{j\in\eta_i}(x_j - \mu) \right]  \right\} = \\ \nonumber & E\left\{ \left[ \sum_{j\in\eta_i}(x_j - \mu) \right] \left[ (x_i - \mu)^3 - 3\beta\sum_{j\in\eta_i}(x_i - \mu)^2 (x_j - \mu) \right. \right. \\ \nonumber \\ \nonumber & \hspace{4cm} \left. \left. + 3\beta^2 \sum_{j\in\eta_i}\sum_{k\in\eta_i}(x_i - \mu)(x_j - \mu)(x_k - \mu) \right. \right. \\ \nonumber  & \hspace{5cm} \left. \left. -\beta^3 \sum_{j\in\eta_i}\sum_{k\in\eta_i}\sum_{l\in\eta_i}(x_j - \mu)(x_k - \mu)(x_l - \mu) \right] \right\}
\end{align}

Thus, by applying the Isserlis' equation to compute the higher-order cross moments as functions of second-order moments, and after some algebraic manipulations, we have:

\begin{align}
	\label{eq:sigma_beta_1}
	I_{\sigma^2 \beta}^{(1)}(\vec{\theta}) & = \frac{1}{\sigma^4}\left[ \sum_{j\in\eta_i}\sigma_{ij} - \beta\sum_{j\in\eta_i}\sum_{k\in\eta_i}\sigma_{jk} \right] - \frac{1}{2\sigma^6}\left[ 6\beta\sum_{j\in\eta_i}\sum_{k\in\eta_i}\sigma_{ij}\sigma_{ik} \right. \\ \nonumber \\ \nonumber & \hspace{3cm} \left. - 3 \beta^2 \sum_{j\in\eta_i}\sum_{k\in\eta_i}\sum_{l\in\eta_i}\left( \sigma_{ij}\sigma_{kl} + \sigma_{ik}\sigma_{jl} + \sigma_{il}\sigma_{jk} \right) \right. \\ \nonumber & \hspace{3cm} \left. + \beta^3 \sum_{j\in\eta_i}\sum_{k\in\eta_i}\sum_{l\in\eta_i}\sum_{m\in\eta_i} \left( \sigma_{jk}\sigma_{lm} + \sigma_{jl}\sigma_{km} + \sigma_{jm}\sigma_{kl} \right) \right]
\end{align}

Moving forward to the next components, it is easy to verify that $I_{\beta\mu}^{(1)}(\vec{\theta}) = I_{\mu\beta}^{(1)}(\vec{\theta}) = 0$ and $I_{\beta\sigma^2}^{(1)}(\vec{\theta}) = I_{\sigma^2\beta}^{(1)}(\vec{\theta})$, since the order of the products in the expectation is irrelevant for the final result. Finally, the last component of the metric tensor $g^{(1)}(\vec{\theta})$ is defined as:

\begin{equation}
	I_{\beta\beta}^{(1)}(\vec{\theta}) = E\left[ \left(\frac{\partial}{\partial\beta} log~p(X; \vec{\theta}) \right)\left(\frac{\partial}{\partial\beta} log~p(X; \vec{\theta}) \right) \right]
\end{equation} which is given by:

\begin{align}
	I_{\beta\beta}^{(1)}(\vec{\theta}) & = \frac{1}{\sigma^4}E\left\{ \left[ (x_i - \mu) - \beta\sum_{j\in\eta_i}(x_j - \mu) \right]^2 \left[ \sum_{j\in\eta_i}(x_j - \mu) \right]^2  \right\} \\ \nonumber & = \frac{1}{\sigma^4} E\left\{ \left[ (x_i - \mu)^2 - 2\beta \sum_{j\in\eta_i} (x_i - \mu)(x_j - \mu) + \beta^2 \sum_{j\in\eta_i}\sum_{k\in\eta_i} (x_j - \mu)(x_k - \mu) \right] \times \right. \\ \nonumber & \left. \hspace{5cm} \left[ \sum_{j\in\eta_i}\sum_{k\in\eta_i} (x_j - \mu)(x_k - \mu) \right] \right\} \\ \nonumber & = \frac{1}{\sigma^4} E \left\{ \sum_{j\in\eta_i}\sum_{k\in\eta_i}(x_i - \mu)(x_i - \mu)(x_j - \mu)(x_k - \mu) \right. \\ \nonumber & \hspace{2cm} \left. - 2\beta\sum_{j\in\eta_i}\sum_{k\in\eta_i} \sum_{l\in\eta_i}(x_i - \mu)(x_j - \mu)(x_k - \mu)(x_l - \mu) \right. \\ \nonumber & \hspace{3cm} \left. + \beta^2 \sum_{j\in\eta_i} \sum_{k\in\eta_i} \sum_{l\in\eta_i} \sum_{m\in\eta_i}(x_j - \mu)(x_k - \mu) (x_l - \mu) (x_m - \mu)  \right\}
\end{align}
 
Using the Isserlis' formula and after some algebra, we have:

\begin{align}
	\label{eq_beta_beta_1}
	I_{\beta\beta}^{(1)}(\vec{\theta}) = \frac{1}{\sigma^2}\sum_{j\in\eta_i} \sum_{k\in\eta_i} \sigma_{jk} & + \frac{1}{\sigma^4} \left[ 2 \sum_{j\in\eta_i} \sum_{k\in\eta_i} \sigma_{ij} \sigma_{ik} \right. \\ \nonumber & \left. - 2\beta \sum_{j\in\eta_i} \sum_{k\in\eta_i} \sum_{l\in\eta_i} \left( \sigma_{ij}\sigma_{kl} + \sigma_{ik}\sigma_{jl} + \sigma_{il}\sigma_{jk} \right) \right. \\ \nonumber & \left. + \beta^2 \sum_{j\in\eta_i} \sum_{k\in\eta_i} \sum_{l\in\eta_i} \sum_{m\in\eta_i} \left( \sigma_{jk}\sigma_{lm} + \sigma_{jl}\sigma_{km} + \sigma_{jm}\sigma_{kl} \right) \right]
\end{align} 

Therefore, we conclude that the type-I Fisher information matrix of an isotropic pairwise Gaussian random field model has the following structure:

\begin{equation}
	g^{(1)}(\vec{\theta}) = \left( \begin{array}{ccc}
	x & 0 & 0 \\ 
	0 & y & w \\ 
	0 & w & z
	\end{array} \right)
\end{equation} where $x = x(\vec{\theta})$, $y = y(\vec{\theta})$, $z = z(\vec{\theta})$ and $w = w(\vec{\theta})$ are the coefficients used to define how we compute an infinitesimal displacement in the manifold (parametric space) around the point $\vec{p} = (\mu, \sigma^2, \beta)$:

\begin{align}
	(ds)^2 & = \begin{bmatrix} d\mu & d\sigma^2 & d\beta \end{bmatrix} \begin{bmatrix} x & 0 & 0 \\ 0 & y & w \\ 0 & w & z \end{bmatrix}\begin{bmatrix} d\mu \\ d\sigma^2 \\ d\beta \end{bmatrix} \\ \nonumber & = x (d\mu)^2 + y (d\sigma^2)^2 + z (d\beta)^2 + 2w(d\beta)(d\sigma^2)
\end{align}

With this we have completely characterized the type-I Fisher information matrix of the isotropic pairwise Gaussian random field model (metric tensor for the parametric space). Note that, from the structure of the Fisher information matrix we see that the parameter $\mu$ is orthogonal to both $\sigma^2$ and $\beta$. In the following, we proceed with the definition of the type-II Fisher information matrix.

\subsubsection{Considerations about the Information Equality}

In the following, we provide a brief discussion based on \cite{Silvey,Casella2002} about the information equality condition, which is a valid property for several probability density function belonging to the exponential family. For purposes of simplification we consider the uniparametric case, knowing that the extension to multiparametric models is quite natural. Let $X$ be a random variable with a probability density function $p(X;\theta)$. Note that:

\begin{equation}
	\frac{\partial^2}{\partial\theta^2} log~p(X;\theta) = \frac{\partial}{\partial\theta}\left[\frac{1}{p(X;\theta)} \frac{\partial}{\partial\theta} p(X;\theta) \right]
\end{equation}

By the product rule we have:

\begin{equation}
	\frac{\partial}{\partial\theta}\left[\frac{1}{p(X;\theta)} \frac{\partial}{\partial\theta} p(X;\theta) \right] = -\frac{1}{p(X;\theta)^2}\left[ \frac{\partial}{\partial\theta}p(X;\theta) \right]^2 + \frac{1}{p(X;\theta)}\frac{\partial^2}{\partial\theta^2}p(X;\theta)
\end{equation} which is leads to

\begin{equation}
	\frac{\partial^2}{\partial\theta^2} log~p(X;\theta) = -\left[ \frac{\partial}{\partial\theta} log~p(X;\theta) \right]^2 + \frac{1}{p(X;\theta)}\frac{\partial^2}{\partial\theta^2}p(X;\theta)
\end{equation}

Rearranging the terms and applying the expectation operator gives us:

\begin{equation}
	E\left[ \left( \frac{\partial}{\partial\theta} log~p(X;\theta) \right)^2 \right] = -E\left[ \frac{\partial^2}{\partial\theta^2} log~p(X;\theta) \right] + E\left[ \frac{1}{p(X;\theta)}\frac{\partial^2}{\partial\theta^2}p(X;\theta) \right]
\end{equation}

By the definition of expected value, the previous expression can be rewritten as:

\begin{equation}
	E\left[ \left( \frac{\partial}{\partial\theta} log~p(X;\theta) \right)^2 \right] = -E\left[ \frac{\partial^2}{\partial\theta^2} log~p(X;\theta) \right] + \int \frac{\partial^2}{\partial\theta^2} p(X;\theta) dx
\end{equation}

Under certain regularity conditions, it is possible to differentiate under the integral sign by interchanging differentiation and integration operators, which implies in:

\begin{equation}
	\int \frac{\partial^2}{\partial\theta^2} p(X;\theta) dx = \frac{\partial^2}{\partial\theta^2} \int p(X;\theta) dx = \frac{\partial^2}{\partial\theta^2} 1 = 0
	\label{eq:integral}
\end{equation} leading to the information equality condition. According to \cite{Silvey}, these regularity conditions can fail for two main reasons: 1) the density function $p(X|\theta)$ may not tail off rapidly enough to ensure the convergence of the integral; 2) the range of integration (the set in $X$ for which $p(X|\theta)$ is non-zero) may depend on the parameter $\theta$. However, note that in the general case the integral defined by equation \eqref{eq:integral} is exactly the difference between the two types of Fisher information, or in a more geometric perspective, between the respective components of the metric tensors $g^{(1)}(\vec{\theta})$ and $g^{(2)}(\vec{\theta})$:

\begin{align}
	\int \frac{\partial^2}{\partial\theta_i \partial\theta_j} p(X;\vec{\theta}) dx & = E\left[ \left( \frac{\partial}{\partial\theta_i} log~p(X;\vec{\theta}) \right) \left( \frac{\partial}{\partial\theta_j} log~p(X;\vec{\theta}) \right)  \right] - \\ \nonumber  & \left\{ - E\left[ \frac{\partial^2}{\partial\theta_i \partial\theta_j} log~p(X;\vec{\theta}) \right] \right\} \\ \nonumber \\ \nonumber & = I_{\theta_i \theta_j}^{(1)}(\vec{\theta}) - I_{\theta_i \theta_j}^{(2)}(\vec{\theta})
\end{align}

We will see in the experiments that these measures (Fisher information), more precisely $I_{\beta \beta}^{(1)}(\vec{\theta})$ and $I_{\beta \beta}^{(2)}(\vec{\theta})$, play an important role in signaling changes in the system's entropy along an evolution of the random field.

\subsubsection{The $g^{(2)}(\vec{\theta})$ Metric Tensor}

By using the second derivative of the log likelihood function, we can compute an alternate metric tensor, given by the type-II Fisher information matrix. The first component of the tensor $g^{(2)}(\vec{\theta})$ is:

\begin{equation}
	I_{\mu\mu}^{(2)} (\vec{\theta}) = -E\left[ \frac{\partial^2}{\partial\mu^2} log~p(X; \vec{\theta}) \right]
\end{equation} which is given by:

\begin{align}
	I_{\mu\mu}^{(2)} (\vec{\theta}) = -\frac{\left( 1 - \beta\Delta \right)}{\sigma^2}E\left\{ \frac{\partial}{\partial\mu}\left[ (x_i - \mu) - \beta\sum_{j\in\eta_i}(x_j - \mu) \right] \right\}	= \frac{1}{\sigma^2}\left( 1 - \beta\Delta \right)^2
\end{align} where $\Delta=8$ is the size of the neighborhood system. The second component is defined by:

\begin{equation}
	I_{\mu\sigma^2}^{(2)} (\vec{\theta}) = -E\left[ \frac{\partial^2}{\partial\mu \partial\sigma^2} log~p(X; \vec{\theta}) \right]
\end{equation} resulting in

\begin{align}
	I_{\mu\sigma^2}^{(2)} (\vec{\theta}) = \frac{1}{\sigma^4}(1 - \beta\Delta) E\left[ (x_i - \mu) - \beta\sum_{j\in\eta_i}(x_j - \mu) \right] = \frac{1}{\sigma^4}(1 - \beta\Delta)\left[ 0 - 0 \right] = 0
\end{align}

Similarly, the third component of the metric tensor is null, since we have:

\begin{align}
	I_{\mu\beta}^{(2)} (\vec{\theta}) & = -E\left[ \frac{\partial^2}{\partial\mu \partial\beta} log~p(X; \vec{\theta}) \right] \\ \nonumber & = \frac{1}{\sigma^2}E\left\{ \Delta \left[ (x_i - \mu) - \beta\sum_{j\in\eta_i}(x_j - \mu) \right] + (1 - \beta\Delta)\left[ \sum_{j\in\eta_i}(x_j - \mu) \right] \right\} \\ \nonumber & = 0 + 0 = 0
\end{align}

Proceeding to the fourth component, it is straightforward to see that $I_\sigma^2\mu(\vec{\theta}) = 0$, since changing the order of the partial derivative operators is irrelevant to the final result. For now, note that both $I_{\mu\mu}^{(2)} (\vec{\theta})$ and $I_{\sigma^2\sigma^2}^{(2)} (\vec{\theta})$ are approximations to $I_{\mu\mu}^{(1)} (\vec{\theta})$ (equation \ref{eq:mu_mu_1}) and $I_{\sigma^2\sigma^2}^{(1)} (\vec{\theta})$ (equation \ref{eq:sigma_sigma_1}) neglecting quadratic and cubic terms of the inverse of the parameter $\sigma^2$, respectively. Thus, we proceed directly to the fifth component, given by:	

\begin{align}
	I_{\sigma^2\sigma^2}^{(2)} (\vec{\theta}) & = - E\left[ \frac{\partial^2}{\partial(\sigma^2)^2} log~p(X; \vec{\theta})  \right] \\ \nonumber & = - E \left\{ \frac{\partial}{\partial\sigma^2}\left[ -\frac{1}{2\sigma^2} + \frac{1}{2\sigma^4} \left( x_i - \mu -\beta\sum_{j\in\eta_i}(x_j - \mu) \right)^2 \right] \right\} \\ \nonumber & = - E \left\{ \frac{1}{2\sigma^4} - \frac{1}{\sigma^6}\left[ (x_i - \mu) - \beta\sum_{j\in\eta_i}(x_j - \mu) \right]^2 \right\} \\ \nonumber & = \frac{1}{2\sigma^4} - \frac{1}{\sigma^6}\left[ 2\beta\sum_{j\in\eta_i} \sigma_{ij} - \beta^2 \sum_{j\in\eta_i}\sum_{k\in\eta_i}\sigma_{jk} \right]
\end{align}

The next component of the metric tensor $g^{(2)}(\vec{\theta})$ is:

\begin{align}
	I_{\sigma^2\beta}^{(2)} (\vec{\theta}) & = - E\left[ \frac{\partial^2}{\partial\sigma^2 \partial\beta} log~p(X; \vec{\theta})  \right] \\ \nonumber & = - E \left\{ \frac{\partial}{\partial\sigma^2}\left[ \frac{1}{\sigma^2} \left( x_i - \mu - \beta\sum_{j\in\eta_i}(x_j - \mu) \right)\left( \sum_{j\in\eta_i}(x_j - \mu) \right) \right]  \right\} \\ \nonumber & = \frac{1}{\sigma^4}\left[ \sum_{j\in\eta_i}\sigma_{ij} - \beta \sum_{j\in\eta_i}\sum_{k\in\eta_i}\sigma_{jk} \right]
\end{align} which is, again, an approximation to $I_{\sigma^2\beta}^{(1)} (\vec{\theta})$ (equation \ref{eq:sigma_beta_1}) obtained by discarding higher-order functions of the parameters $\sigma^2$ and $\beta$. It is straightforward to see that the next two components of $g^{(2)}(\vec{\theta})$ are identical to their symmetric counterparts, that is, $I_{\beta\mu}^{(2)} (\vec{\theta}) = I_{\mu\beta}^{(2)} (\vec{\theta}) = 0$ and $I_{\beta\sigma^2}^{(2)} (\vec{\theta}) = I_{\sigma^2\beta}^{(2)} (\vec{\theta})$. Finally, we have the last component of the Fisher information matrix:

\begin{align}
	I_{\beta\beta}^{(2)} (\vec{\theta}) = - E\left[ \frac{\partial^2}{\partial\beta^2} log~p(X; \vec{\theta}) \right]
\end{align} which is given by:

\begin{align}
	I_{\beta\beta}^{(2)} (\vec{\theta}) & = -\frac{1}{\sigma^2}E \left\{\frac{\partial}{\partial\beta}\left[\left(  x_i - \mu - \beta\sum_{j\in\eta_i}(x_j - \mu)\right) \left( \sum_{j\in\eta_i}(x_j - \mu) \right) \right] \right\} \\ \nonumber & = \frac{1}{\sigma^2}E\left[ \left( \sum_{j\in\eta_i}(x_j - \mu) \right) \left( \sum_{j\in\eta_i}(x_j - \mu) \right) \right] \\ \nonumber & = \frac{1}{\sigma^2}\sum_{j\in\eta_i}\sum_{k\in\eta_i}\sigma_{jk}
\end{align}

Once again, note that $I_{\beta\beta}^{(2)} (\vec{\theta})$ is an approximation to $I_{\beta\beta}^{(1)} (\vec{\theta})$ (equation \ref{eq_beta_beta_1}) where higher-order functions of the parameters $\sigma^2$ and $\beta$ are suppressed. It is clear that the difference between the components of the two metric tensors $g^{(1)}(\vec{\theta})$ and $g^{(2)}(\vec{\theta})$ is significant when the inverse temperature parameter is not null. On the other hand, the global structure of $g^{(2)}(\vec{\theta})$ is essentially the same of $g^{(1)}(\vec{\theta})$, implying that the definition of $ds^2$ is identical to the previous case, but with different coefficients for $(d\mu)^2$, $(d\sigma^2)^2$, $(d\beta)^2$ and $(d\beta)(d\sigma^2)$. Note also that when the inverse temperature parameter is fixed at zero, both metric tensors converge to:

\begin{equation}
	g^{(0)}(\vec{\theta}) = \left( \begin{array}{ccc}
	\frac{1}{\sigma^2} & 0 & 0 \\ 
	0 & \frac{1}{2\sigma^4} & 0 \\ 
	0 & 0 & \Delta
	\end{array} \right)
\end{equation} where $\Delta=8$ is a constant defining the support of the neighborhood system. This is exactly the Fisher information matrix of a traditional Gaussian random variable $X \sim N(\mu, \sigma^2)$ (excluding the third row and column), as it would be expected.

\subsubsection{Expressing Fisher information in tensorial notation}

\begin{figure}[h]
\begin{center}
\includegraphics[scale=0.4]{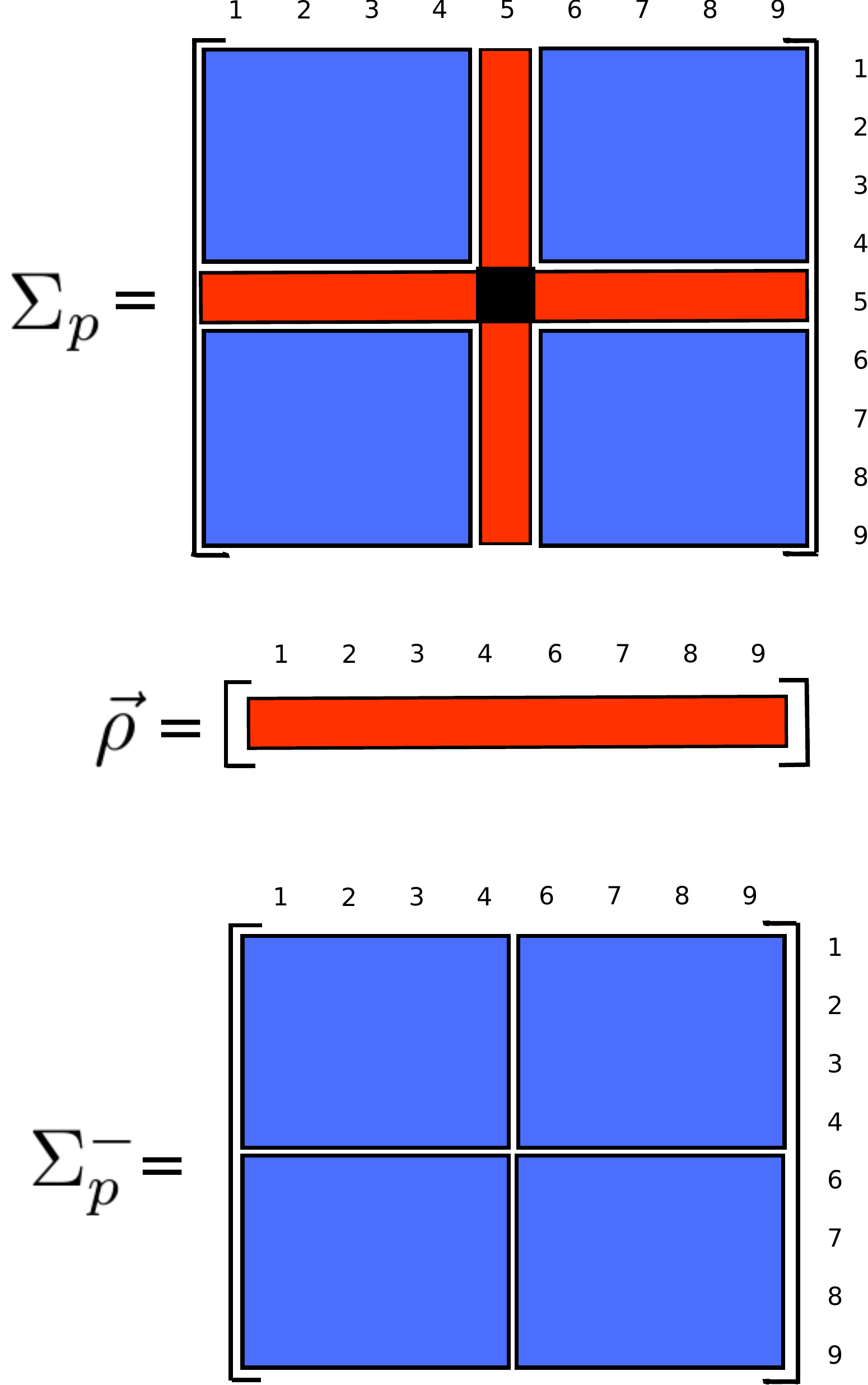}
\end{center}
\caption{{\bf Decomposing the covariance matrix $\Sigma_{p}$ into $\Sigma_{p}^{-}$ and $\vec{\rho}$ on a second-order neighborhood system ($\Delta=8$).} 
By expressing the components of the metric tensors in terms of Kronocker products, it is possible to compute Fisher information in a efficient way during computational simulations.
}
\label{fig:cov_matrix}
\end{figure}

In order to simplify the notations and also to make computations faster, the expressions for the components of the metric tensors $g^{(1)}(\vec{\theta})$ and $g^{(2)}(\vec{\theta})$ can be rewritten in a matrix-vector form using a tensor notation. Let $\Sigma_{p}$ be the covariance matrix of the random vectors $\vec{p}_{i}, i = 1,2,\ldots,n$, obtained by lexicographic ordering the local configuration patterns $x_{i} \cup \eta_{i}$ for a snapshot of the system (a static configuration $\mathbf{X}^{(t)}$). In this work, we choose a second-order neighborhood system, making each local configuration pattern a $3 \times 3$ patch. Thus, since each vector $\vec{p}_{i}$ has 9 elements, the resulting covariance matrix $\Sigma_{p}$ is $9 \times 9$. Let $\Sigma_{p}^{-}$ be the sub-matrix of dimensions $8 \times 8$ obtained by removing the central row and central column of $\Sigma_{p}$ (these elements are the covariances between the central variable $x_{i}$ and each one of its neighbors $x_{j} \in \eta_{i}$). Also, let $\vec{\rho}$~be the vector of dimensions $8 \times 1$ formed by all the elements of the central row of $\Sigma_{p}$, excluding the middle one (which denotes the variance of $x_{i}$ actually). Fig. \ref{fig:cov_matrix} illustrates the process of decomposing the covariance matrix $\Sigma_{p}$ into the sub-matrix $\Sigma_{p}^{-}$ and the vector $\vec{\rho}$~ in an isotropic pairwise GMRF model defined on a second-order neighborhood system (considering the 8 nearest neighbors).

Given the above, we can express the elements of the Fisher information matrix in a tensorial form using Kronecker products. The following definitions provide a computationally efficient way to numerically evaluate $g^{(1)}(\vec{\theta})$ exploring tensor products.

\begin{mydef}
Let an isotropic pairwise Gaussian Markov random field be defined on a lattice $S=\left\{ s_{1}, s_{2}, \ldots , s_{n} \right\}$ with a neighborhood system $\eta_{i}$ of size $\Delta$ (usual choices for $\Delta$ are even values: 4, 8, 12, 20, 24,...). Assuming that the set $\mathbf{X^{(t)}}=\{x_{1}^{(t)}, x_{2}^{(t)}, \ldots, x_{n}^{(t)} \}$ denotes the global configuration of the system at iteration $t$, and both $\vec{\rho}$ and $\Sigma_{p}^{-}$ are defined according to Figure \ref{fig:cov_matrix}, the components of the metric tensor $g^{(1)}(\vec{\theta})$ (Fisher information matrix) can be expressed as: 
\end{mydef}  

\begin{equation}
	I_{\mu\mu}^{(1)}(\vec{\theta}) = \frac{1}{\sigma^2}\left(1-\beta \Delta \right)^2\left[ 1 - \frac{1}{\sigma^2}\left( 2\beta\left\| \vec{\rho} \right\|_{+} - \beta^2 \left\| \Sigma_{p}^{-} \right\|_{+} \right) \right] 
\end{equation}

\begin{align}	
	I_{\sigma^2\sigma^2}^{(1)}(\vec{\theta}) = \frac{1}{2\sigma^4} & - \frac{1}{\sigma^6} \left[ 2\beta\left\| \vec{\rho} \right\|_{+} - \beta^2 \left\| \Sigma_{p}^{-} \right\|_{+} \right] \\ \nonumber & + \frac{1}{\sigma^8}\left[ 3\beta^2 \left\| \vec{\rho} \otimes \vec{\rho} \right\|_{+} - 3 \beta^3 \left\| \vec{\rho} \otimes \Sigma_{p}^{-} \right\|_{+} + 3\beta^4 \left\| \Sigma_{p}^{-} \otimes \Sigma_{p}^{-} \right\|_{+}  \right] \nonumber
\end{align}	
	
\begin{align}
	I_{\sigma^2\beta}^{(1)}(\vec{\theta}) = I_{\beta\sigma^2}^{(1)}(\vec{\theta}) & = \frac{1}{\sigma^4}\left[ \left\| \vec{\rho} \right\|_{+} - \beta \left\| \Sigma_{p}^{-} \right\| \right] \\ \nonumber & - \frac{1}{2\sigma^6} \left[ 6\beta \left\| \vec{\rho} \otimes \vec{\rho} \right\|_{+} - 9 \beta^2 \left\| \vec{\rho} \otimes \Sigma_{p}^{-} \right\|_{+} + 3\beta^3 \left\| \Sigma_{p}^{-} \otimes \Sigma_{p}^{-} \right\|_{+}  \right] \nonumber
\end{align}	

\begin{equation}	
	I_{\beta\beta}^{(1)}(\vec{\theta}) = \frac{1}{\sigma^2} \left\| \Sigma_{p}^{-} \right\|_{+} + \frac{1}{\sigma^4} \left[ 2 \left\| \vec{\rho} \otimes \vec{\rho} \right\|_{+} - 6 \beta \left\| \vec{\rho} \otimes \Sigma_{p}^{-} \right\|_{+} + 3\beta^2 \left\| \Sigma_{p}^{-} \otimes \Sigma_{p}^{-} \right\|_{+}  \right] 
\end{equation} where $\left\| A \right\|_{+}$ denotes the summation of all the entries of the vector/matrix $A$ (not to be confused with the norm) and $\otimes$ denotes the Kronecker (tensor) product. Similarly, we can express the components of the metric tensor $g^{(2)}(\vec{\theta})$ in this form.

\begin{mydef}
Let an isotropic pairwise Gaussian Markov random field be defined on a lattice $S=\left\{ s_{1}, s_{2}, \ldots , s_{n} \right\}$ with a neighborhood system $\eta_{i}$ of size $\Delta$ (usual choices for $\Delta$ are even values: 4, 8, 12, 20, 24,...). Assuming that the set $\mathbf{X^{(t)}}=\{x_{1}^{(t)}, x_{2}^{(t)}, \ldots, x_{n}^{(t)} \}$ denotes the global configuration of the system at iteration $t$, and both $\vec{\rho}$ and $\Sigma_{p}^{-}$ are defined according to Figure \ref{fig:cov_matrix}, the components of the metric tensor $g^{(2)}(\vec{\theta})$ (Fisher information matrix) can be expressed as: 
\end{mydef}

\begin{equation}
	I_{\mu\mu}^{(2)}(\vec{\theta}) = \frac{1}{\sigma^2}\left(1-\beta \Delta \right)^2
\end{equation}	

\begin{equation}	
	I_{\sigma^2\sigma^2}^{(2)}(\vec{\theta}) = \frac{1}{2\sigma^4} - \frac{1}{\sigma^6} \left[ 2\beta\left\| \vec{\rho} \right\|_{+} - \beta^2 \left\| \Sigma_{p}^{-} \right\|_{+} \right] 
\end{equation}	
	
\begin{equation}
	I_{\sigma^2\beta}^{(2)}(\vec{\theta}) = I_{\beta\sigma^2}^{(1)}(\vec{\theta}) = \frac{1}{\sigma^4}\left[ \left\| \vec{\rho} \right\|_{+} - \beta \left\| \Sigma_{p}^{-} \right\| \right] 
\end{equation}	

\begin{equation}	
	I_{\beta\beta}^{(2)}(\vec{\theta}) = \frac{1}{\sigma^2} \left\| \Sigma_{p}^{-} \right\|_{+}  
\end{equation}

From the above equations it is clear to see that the components of $g^{(2)}(\vec{\theta})$ are approximations to the components of $g^{(1)}(\vec{\theta})$, obtained by discarding the higher-order terms (the cross Kronecker products vanish).

\subsection{Entropy in Gaussian random fields}

Entropy is one of the most ubiquitous concepts in science, with applications in a large number of research fields. In information theory, Shannon entropy is the most widely know statistical measure related to a random variable, since it often characterizes a degree of uncertainty about any source of information \cite{shannon1949}. Similarly, in statistical physics, entropy plays an important role in thermodynamics, being a relevant measure in the the study and analysis of complex dynamical systems \cite{jaynes1957}. In this paper, we try to understand entropy in a more geometrical perspective, by means of its relation to Fisher information.

Our definition of entropy in a Gaussian random field is done by repeating the same process employed to derive the Fisher information matrices. Knowing that the entropy of random variable x is defined by the expected value of self-information, given by $-log~p(x)$, we have the following definition.

\begin{mydef}
Let a pairwise GMRF be defined on a lattice $S=\left\{ s_{1}, s_{2}, \ldots , s_{n} \right\}$ with a neighborhood system $\eta_{i}$. Assuming that the set of observations $\mathbf{X^{(t)}}=\{x_{1}^{(t)}, x_{2}^{(t)}, \ldots, x_{n}^{(t)} \}$ denote the global configuration of the system at time $t$, then the entropy $H_{\beta}$ for this state $\mathbf{X^{(t)}}$ is given by:
\end{mydef}

\begin{align}
	\label{eq:entropia1}
	 H_{\beta} = - E\left[ log~p\left(x_{i}| \eta_{i}, \vec{\theta} \right) \right] & = \frac{1}{2}\left[ log\left( 2\pi\sigma^2 \right) + 1\right] \\ \nonumber & - \frac{1}{\sigma^2} \left[ \beta\sum_{j \in \eta_i}\sigma_{ij} - \frac{\beta^2}{2}\sum_{j \in \eta_i}\sum_{k \in \eta_i}\sigma_{jk} \right]
\end{align}

Note that, for $\beta=0$ the expression is reduced to the entropy of a simple Gaussian random variable, as it would be expected. By using the tensor notation, we have:

\begin{align}     
	H_{\beta} = H_{G} - \frac{1}{\sigma^2}\left[ \beta\left\| \vec{\rho} \right\|_{+} - \frac{\beta^{2}}{2}\left\| \Sigma_{p}^{-}\right\|_{+} \right] = H_{G} - \left[ \frac{\beta}{\sigma^{2}}\left\| \vec{\rho} \right\|_{+} - \frac{\beta^{2}}{2} I_{\beta\beta}^{(2)}(\vec{\theta}) \right]
	\label{eq:entropy}
\end{align} where $H_{G}$ denotes the entropy of a Gaussian random variable with mean $\mu$ and variance $\sigma^2$, and $I_{\beta\beta}^{(2)}(\vec{\theta})$ is a component of the Fisher information matrix $g^{(2)}(\vec{\theta})$. In other words, entropy is related to Fisher information. We will see in the experimental results that the analysis of Fisher information can bring us insights in predicting whether the entropy of the system is increasing or decreasing.

\subsection{Maximum Pseudo-Likelihood Estimation}

A fundamental step in our simulations is the computation of the Fisher information matrix (metric tensor components) and entropy, given an output of the random field model. All these measures are function of the model parameters, more precisely, of the variance and the inverse temperature. In all the experiments conducted in this investigation, the Gaussian random field parameters $\mu$ and $\sigma^2$ are both estimated by the sample mean and variance, respectively, using the maximum likelihood estimatives. However, maximum likelihood estimation is intractable for the inverse temperature parameter estimation ($\beta$), due to the existence of the partition function in the joint Gibbs distribution. An alternative, proposed by Besag \cite{Besag1974}, is to perform maximum pseudo-likelihood estimation, which is based on the conditional independence principle. The basic idea with this proposal is to replace the independence assumption by a more flexible conditional independence hypothesis, allowing us to use the local conditional density functions of the random field model in the definition of a likelihood function, called pseudo-likelihood.

 It has been shown that maximum likelihood estimators are asymptotically efficient, that is, the uncertainty in the estimation of unknown parameters is minimized. In order to quantify the uncertainty in the estimation of the inverse temperature parameter, it is necessary to compute the asymptotic variance of the maximum pseudo-likelihood estimator. We will see later that the components $I_{\beta\beta}^{(1)}(\vec{\theta})$ and $I_{\beta\beta}^{(2)}(\vec{\theta})$ of both tensors $g^{(1)}(\vec{\theta})$ and $g^{(2)}(\vec{\theta})$ are crucial in quantifying this uncertainty. First, we need to define the pseudo-likelihood function of a random field model.
 
 \begin{mydef}
Let an isotropic pairwise Markov random field model be defined on a rectangular lattice $S=\left\{ s_{1}, s_{2}, \ldots , s_{n} \right\}$ with a neighborhood system $\eta_{i}$. Assuming that $\mathbf{X^{(t)}}=\{x_{1}^{(t)}, x_{2}^{(t)}, \ldots, x_{n}^{(t)} \}$ denotes the set corresponding to the observations at a time $t$ (a snapshot of the random field), the pseudo-likelihood function of the model is defined by:
\end{mydef}

\begin{equation}
	L\left(\vec{\theta}; \mathbf{X}^{(t)}\right) = \prod_{i=1}^{n}p( x_{i} | \eta_{i}, \vec{\theta} )
	\label{eq:PL}
\end{equation} where $\vec{\theta} = (\mu, \sigma^2, \beta)$. The pseudo-likelihood function is the product of the local conditional density functions throughout the field viewed as a function of the model parameters. For an isotropic pairwise Gaussian Markov random field, the pseudo-likelihood function is given by plugging equation \eqref{eq:GMRF} into equation \eqref{eq:PL}:

\begin{equation}
	log~L\left(\vec{\theta}; \mathbf{X}^{(t)} \right) = -\frac{n}{2}log\left( 2\pi\sigma^{2} \right) -\frac{1}{2\sigma^{2}}\sum_{i=1}^{n}\left[ x_{i} - \mu - \beta\sum_{j \in \eta_i}\left( x_{j} - \mu \right) \right]^{2}
	\label{eq:GMRF_PL} 
\end{equation}

By differentiating equation \eqref{eq:GMRF_PL} with respect to $\beta$ and properly solving the pseudo-likelihood equation, we obtain the following estimator for the inverse temperature parameter:

\begin{equation}
	\hat{\beta}_{MPL} = \frac{\displaystyle\sum_{i=1}^{n}\left[\left( x_{i} - \mu \right)\displaystyle\sum_{j \in \eta_i}\left(x_{j} - \mu \right)\right]}{\displaystyle\sum_{i=1}^{n}\left[ \displaystyle\sum_{j \in \eta_i}\left( x_{j} - \mu  \right) \right]^{2}}
	\label{eq:BetaMPL}
\end{equation}

Assuming that the random field is defined on a retangular 2D lattice where the cardinality of the neighborhood system is fixed ($\Delta$), the maximum pseudo-likelihood estimator for the inverse temperature parameter can be rewritten as:

\begin{equation}
	\hat{\beta}_{MPL} = \frac{\displaystyle\sum_{j \in \eta_i}{\sigma}_{ij}}{\displaystyle\sum_{j \in \eta_i}\displaystyle\sum_{k \in \eta_i}{\sigma}_{jk}} = \frac{\left\| \vec{\rho} \right\|_{+}}{\left\| \Sigma_{p}^{-}\right\|_{+}}
	\label{eq:BetaMPL2}
\end{equation} which means that we can also compute this estimative from the covariance matrix of the configuration patterns. In other words, given a snapshot of the system at an instant $t$, $\mathbf{X^{(t)}}$, all the measures we need are based solely in the matrix $\Sigma_{p}$. Therefore, in terms of information geometry, a sequence of Gaussian random field outputs in time can be summarized into a sequence of covariance matrices. In computational terms, it means a huge reduction in the volume of data.

In our computational simulations, we fix initial values for the parameters $\mu$, $\sigma^2$ and $\beta$, and at each iteration an infinitesimal displacement in the inverse temperature ($\beta$ axis) is performed. A new random field output is generated for each iteration and in order to avoid any degree of supervision throughout the process of computing the entropy and both Fisher information metrics of each configuration, the unknown model parameters are properly estimated from data. 

However, in estimating the inverse temperature parameter of random fields via maximum pseudo-likelihood, a relevant question emerges: how to measure the uncertainty in the estimation of $\beta$? Is it possible to quantify this uncertainty? We will see that both versions of Fisher information play a central role in answering this question.

\subsection{Uncertainty in the Estimation of the Inverse Temperature}

It is known from the statistical inference literature that both maximum likelihood and maximum pseudo-likelihood estimators share an important property: asymptotic normality \cite{Jensen,winkler}. It is possible, therefore, to characterize their behavior in the limiting case by knowing the asymptotic variance. A limitation from maximum pseudo-likelihood approach is that there is no result proving that this method is asymptotically efficient (maximum likelihood estimators have been shown to be asymptotically efficient since in the limiting case their variance reaches the Cramer-Rao lower bound). It is known that the asymptotic variance of the inverse temperature parameter in an isotropic pairwise GMRF is given by \cite{Levada2014}:

\begin{equation}
	\upsilon_{\beta} = \frac{I_{\beta\beta}^{(1)}(\vec{\theta})}{[I_{\beta\beta}^{2}(\vec{\theta})]^2} = \frac{1}{I_{\beta\beta}^{(2)}(\vec{\theta})} + \frac{1}{I_{\beta\beta}^{(2)}(\vec{\theta})^{2}}\left(I_{\beta\beta}^{(1)}(\vec{\theta}) - I_{\beta\beta}^{(2)}(\vec{\theta}) \right)
\end{equation} showing that in the information equilibrium condition, that is, $I_{\beta\beta}^{(1)}(\vec{\theta}) = I_{\beta\beta}^{(2)}(\vec{\theta})$, we have the traditional Cramer-Rao lower bound, given by the inverse of the Fisher information.

A very simple interpretation of this equation indicates that the uncertainty in the estimation of the inverse temperature parameter is reduced when $I_{\beta\beta}^{(1)}(\vec{\theta})$ is minimized and $I_{\beta\beta}^{(2)}(\vec{\theta})$ is maximized. Essentially, it means that most local patterns must be aligned to the expected global behavior and, in average, the local likelihood functions should not be flat (indicating that there is a small number of candidates for $\beta$).

\subsection{Fisher Curves and the Information Space}

By computing $I_{\theta_i\theta_j}^{(1)}(\vec{\theta})$, $I_{\theta_i\theta_j}^{(2)}(\vec{\theta})$ and $H_{\beta}$, we have access to three important information theoretic measures regarding a global configuration $\mathbf{X}^{(t)}$ of the random field. We call the 3D space generated by these 3 measures, the information space. A point in this space represents the value of that specific component of the metric tensor, $I_{\theta_i\theta_j}(\vec{\theta})$, when the system's entropy value is $H(\beta)$. This allows us to define the Fisher curves of the system.

\begin{mydef}
Let an isotropic pairwise GMRF model be defined on a lattice $S=\left\{ s_{1}, s_{2}, \ldots , s_{n} \right\}$ with a neighborhood system $\eta_{i}$ and $\mathbf{X}^{(\beta_{1})},\mathbf{X}^{(\beta_{2})},\ldots,\mathbf{X}^{(\beta_{n})}$ be a sequence of outcomes (global configurations) produced by different values of $\beta_{i}$ (inverse temperature parameters) for which $A = \beta_{MIN} = \beta_{1} < \beta_{2} < \cdots < \beta_{n} = \beta_{MAX} = B$. The Fisher curve from $A$ to $B$ is defined as the parametric curve $\vec{F}:\Re \rightarrow \Re^{3}$ that maps each configuration $\mathbf{X}^{(\beta_{i})}$ to a point $\left(I_{\theta_i\theta_j}^{(1)}(\beta) , I_{\theta_i\theta_j}^{(2)}(\beta), H(\beta) \right)$ in the information space:
\end{mydef}

\begin{equation}
	\vec{F}_{A}^{B}\left(\beta \right) = \left(I_{\theta_i\theta_j}^{(1)}(\beta) , I_{\theta_i\theta_j}^{(2)}(\beta), H(\beta) \right) \qquad\qquad \beta = A,\ldots,B
\end{equation} where $I_{\theta_i\theta_j}^{(1)}(\beta)$ and $I_{\theta_i\theta_j}^{(2)}(\beta)$ denote the $(i,j)$ components of the metric tensors $g^{(1)}(\vec{\theta})$ and $g^{(2)}(\vec{\theta})$, respectively, and $H(\beta)$ denotes the entropy.

The motivation behind the Fisher curve is the development of a computational tool for the study and characterization of random fields. Basically, the Fisher curve of a system is the parametric curve embedded in this information-theoretic space obtained by varying the inverse temperature parameter $\beta$ from an initial value $\beta_I$ to a final value $\beta_F$. The resulting curve provides a geometrical interpretation about how the random field evolves from a lower entropy configuration A to a higher entropy configuration B (or vice-versa), since the Fisher information plays an important role in providing a natural metric to the Riemannian manifold of a statistical model \cite{Amari2000,Kass1989}. We will call the path from a global system configuration A to a global system configuration B as the \emph{Fisher curve} (from A to B) of the system, denoted by $\vec{F}_{A}^{B}(\beta)$. Instead of using the notion of time as parameter to build the curve $\vec{F}$, we parametrize $\vec{F}$ by the inverse temperature parameter $\beta$. In geometrical terms, we are trying to measure the deformation in the metric tensor of the stochastic model (local geometric property) induced by a displacement in the inverse temperature parameter direction.

We are especially interested in characterizing random fields by measuring and quantifying their behavior as the inverse temperature parameter deviates from zero, that is, when temperature leaves infinity. As mentioned before, the isotropic pairwise GMRF model belongs to the regular exponential family of distributions when the inverse temperature parameter is zero ($T = \infty$). In this case, it has been shown that the geometric structure, whose natural Riemannian metric is given by the Fisher information matrix (metric tensor), has constant negative curvature (hyperbolic geometry). Besides, Fisher information can be measured by two different but equivalent ways (information equality). 

As the inverse temperature increases, the model starts to deviate from this known scenario, and the original Riemannian metric does not correctly represents the geometric structure anymore (since there is an additional parameter). The manifold which used to be 2D (surface) now slowly is transformed (deformed) to a different structure. In other words, as this extra dimension is gradually emerging (since $\beta$ not null), the metric tensor is transformed (the original $2 \times 2$ Fisher information matrix becomes a $3 \times 3$ matrix). We believe that the intrinsic notion of time in the evolution of a random field composed by Gaussian variables is caused by the irreversibility of this deformation process, as the results suggest.

\section{Computational Simulations}

In this section, we present some experimental results using computational methods for simulating the dynamics and evolution of Gaussian random fields. All the simulations were performed by applying Markov Chain Monte Carlo (MCMC) algorithms for the generation of random field outcomes based on the specification of the model parameters. In this paper, we make intensive use of the Metropolis-Hastings algorithm \cite{Metropolis1953}, a classic method in the literature. All the computational implementations are done using the Python Anaconda platform, which includes several auxiliary packages for scientific computing.

The main objective here is to measure $I_{\theta_i\theta_j}^{(1)}(\beta)$, $I_{\theta_i\theta_j}^{(2)}(\beta)$ and $H(\beta)$ along a MCMC simulation in which the inverse temperature parameter $\beta$ is controlled to guide the global system behavior. Initially, $\beta$ is set to $\beta_{MIN}=0$, that is, the initial temperature is infinite. In the following, $\beta$ is linearly increased, with fixed increments $\Delta\beta$, up to an upper limit $\beta_{MAX}$. After that, the exact reverse process is performed, that is, the inverse temperature $\beta$ is linearly decreased using the same fixed increments ($-\Delta\beta$) all the way down to zero. With this procedure, we are actually performing a positive displacement followed by a negative displacement along the inverse temperature parameter ``direction'' in the parametric space. By sensing each component of the metric tensor (Fisher information) at each point, we are essentially trying to capture the deformation in the geometric structure of the statistical manifold (parametric space) throughout the process. 

\begin{figure}[h]
\begin{center}
\includegraphics[scale=0.6]{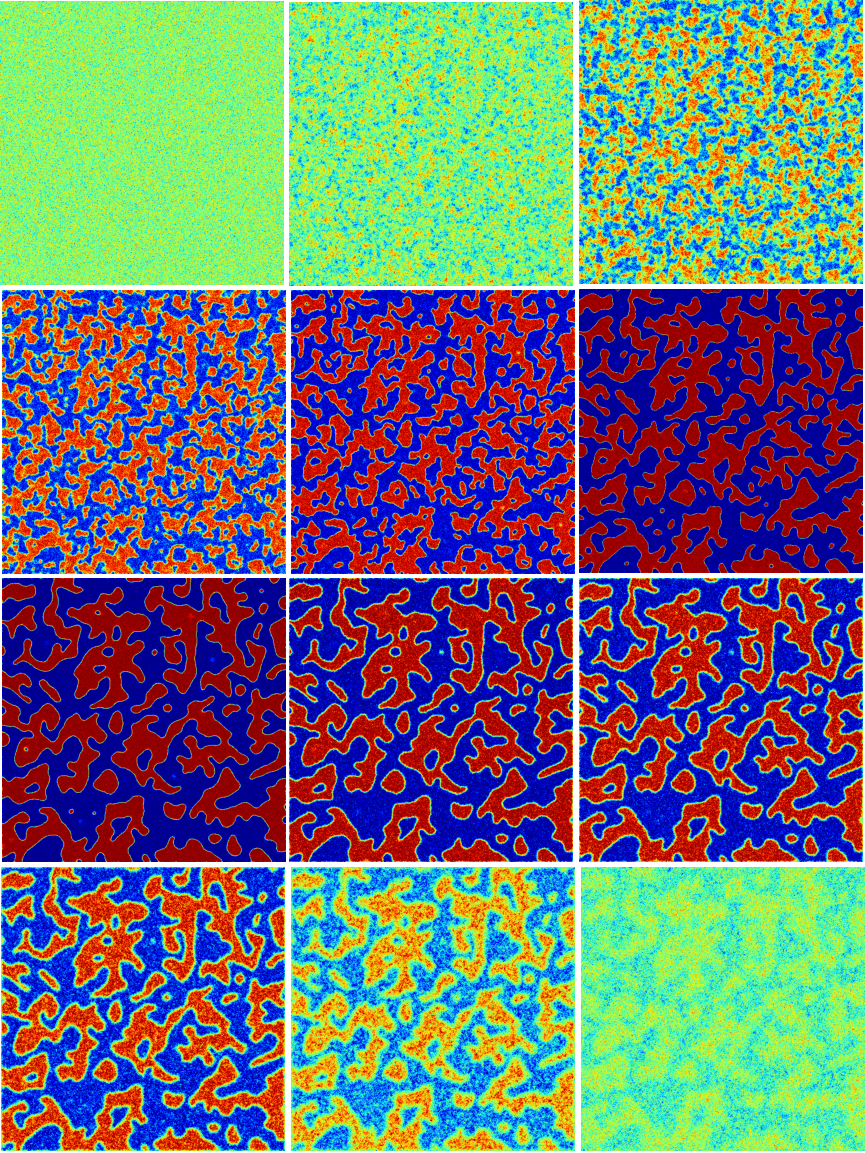}
\end{center}
\caption{{\bf Random field dynamics along a Markov Chain Monte Carlo (MCMC) simulation.} 
Evolution of the random field as the inverse temperature parameter $\beta$ is first increased from zero to 0.5 and then decreased from 0.5 to zero.
}
\label{fig:GMRF_configs}
\end{figure}

The simulations were performed using the following parameter settings: $\mu = 0$, $\sigma^{2} = 5$ (initial value), $A=\beta_{MIN}=0$, $B=\beta_{MAX}=0.5$, $\Delta\beta = 0.001$ and 1000 iterations. At the end of a single MCMC simulation, 2.1 GB of data is generated, representing 1000 random field configurations of size $512 \times 512$. Fig. \ref{fig:GMRF_configs} shows some samples of the random field during the evolution of the system.

\subsection{Fisher Curves in Gaussian Random Fields}

The goal of this investigation is to analyse the behavior of the metric tensor of the statistical manifold of a Gaussian random field by learning everything from data, including the inverse temperature parameter $\beta$. At each iteration of the simulation, the values of $\mu$ and $\sigma^{2}$ are updated by computing the sample mean and sample variance, respectively. The inverse temperature parameter is updated by computing the maximum pseudo-likelihood estimative. 

\begin{figure}[h]
\begin{center}
\includegraphics[scale=0.4]{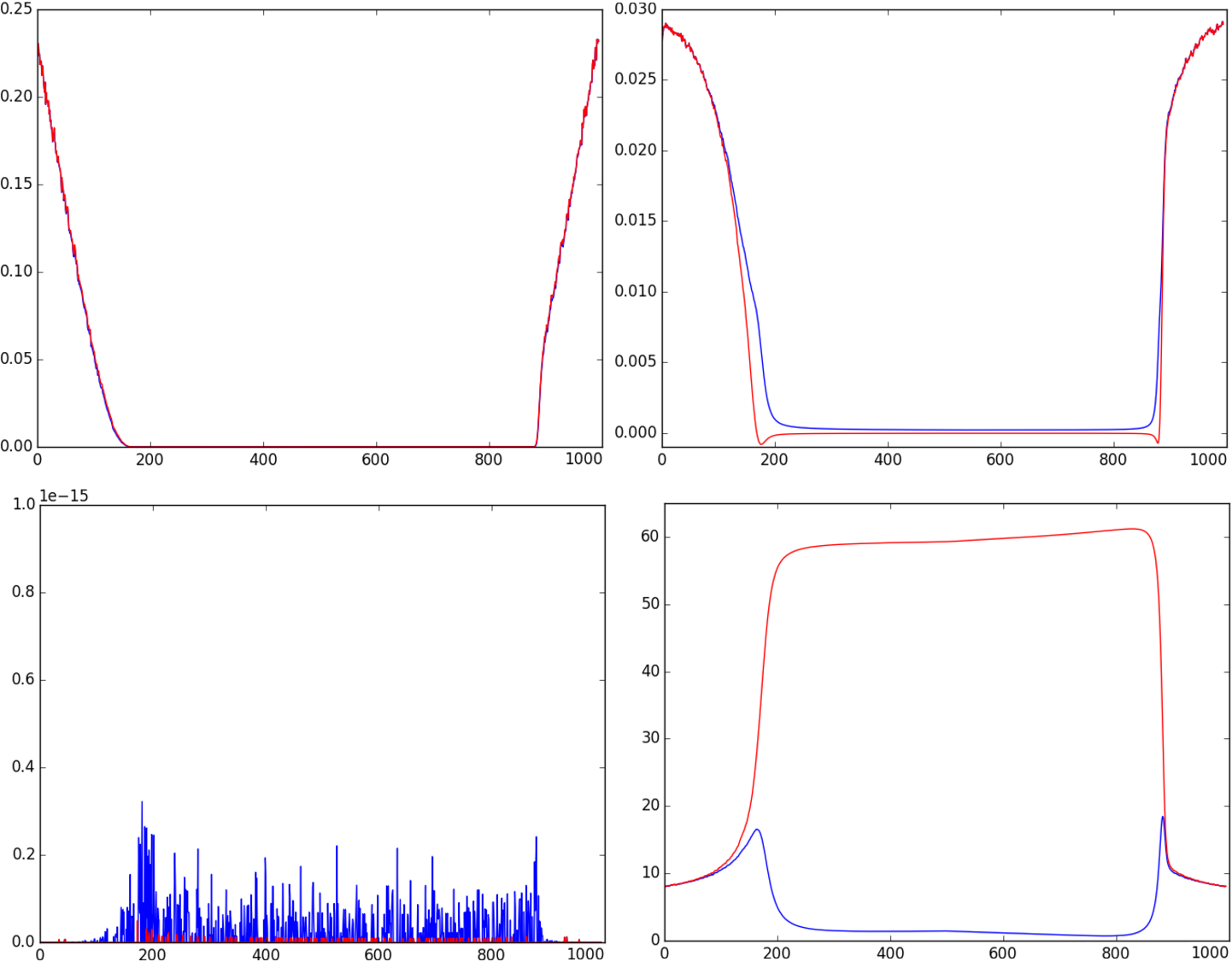}
\end{center}
\caption{{\bf Variation in the components of the metric tensor of the statistical manifold (Fisher information) along the MCMC simulation.}
The blue lines represent the components of the $g^{(1)}(\vec{\theta})$ tensor and the red lines represent the components of the $g^{(2)}(\vec{\theta})$ tensor. The first row shows the graphs of  $I_{\mu\mu}^{(1)}(\vec{\theta})$ versus $I_{\mu\mu}^{(2)}(\vec{\theta})$ and $I_{\sigma^{2}\sigma^{2}}^{(1)}(\vec{\theta})$ versus $I_{\sigma^{2}\sigma^{2}}^{(2)}(\vec{\theta})$. The second row shows the graphs of $I_{\sigma^{2}\beta}^{(1)}(\vec{\theta})$ versus $I_{\sigma^{2}\beta}^{(2)}(\vec{\theta})$ and  $I_{\beta\beta}^{(1)}(\vec{\theta})$ versus $I_{\beta\beta}^{(2)}(\vec{\theta})$. Note that, from an information geometry perspective, the most relevant component in this geometric deformation process of the statistical manifold is the one regarding the inverse temperature parameter. Two important aspects that must be remarked are: 1) there is a large divergence between $I_{\beta\beta}^{(1)}(\vec{\theta})$ and $I_{\beta\beta}^{(2)}(\vec{\theta})$, that is, the information equality condition fails when $\beta$ deviates from zero; 2) Although the total displacement in the $\beta$ ``axis'' adds up to zero, $I_{\beta\beta}^{(1)}(\vec{\theta})$ is highly asymmetric, which indicates that the deformations induced by the metric tensor to the statistical manifold when entropy is increasing are different from those when entropy is decreasing.
}
\label{fig:Fisher}
\end{figure}

In order to sense the local geometry of the parametric space during the random field dynamics, we have computed the values of all the components of the metric tensor at each iteration of the simulation. Since we are dealing with both forms of Fisher information (using the square of the first derivative and the negative of the second derivative) to investigate the information equality condition, both $g^{(1)}(\vec{\theta})$ and $g^{(2)}(\vec{\theta})$ tensors are being estimated. Fig. \ref{fig:Fisher} shows a comparison between each component of $g^{(1)}(\vec{\theta})$ with its corresponding component in $g^{(2)}(\vec{\theta})$ along the entire simulation. At this point, some important aspects must be discussed. First, these results show that the components $I_{\mu\mu}(\vec{\theta})$, $I_{\sigma^{2}\sigma^{2}}(\vec{\theta})$ and $I_{\sigma^{2}\beta}(\vec{\theta})$ are practically negligible in comparison to $I_{\beta\beta}(\vec{\theta})$ in terms of magnitude. Second, while the differences $I_{\mu\mu}^{(1)}(\vec{\theta}) - I_{\mu\mu}^{(2)}(\vec{\theta})$,  $I_{\sigma^{2}\sigma^{2}}^{(1)}(\vec{\theta}) - I_{\sigma^{2}\sigma^{2}}^{(2)}(\vec{\theta})$ and $I_{\sigma^{2}\beta}^{(1)}(\vec{\theta}) - I_{\sigma^{2}\beta}^{(2)}(\vec{\theta})$ are also negligible, the difference $I_{\beta\beta}^{(1)}(\vec{\theta}) - I_{\beta\beta}^{(2)}(\vec{\theta})$ is very significant, especially for larger values of $\beta$. And third, note that even though the total displacement in the inverse temperature direction adds up to zero (since $\beta$ is updated from zero to 0.5 and back), $I_{\beta\beta}(\vec{\theta})$ is highly asymmetric, which indicates that the deformations induced by the metric tensor to the statistical manifold when entropy is increasing are different than those when entropy is decreasing.

\begin{figure}[h]
\begin{center}
\includegraphics[scale=1]{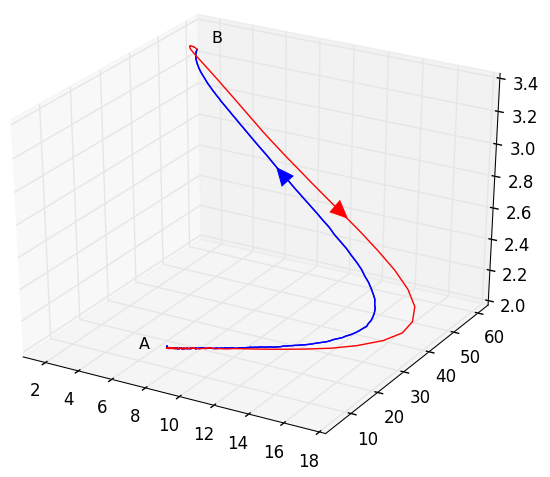}
\end{center}
\caption{{\bf Fisher curve of the random field regarding the component $I_{\beta\beta}(\vec{\theta})$.} 
The parametric curve was built by varying the inverse temperature parameter $\beta$ from $\beta_{MIN}=0$ (state A) to $\beta_{MAX}=0.5$ (state B) and back. The results show that moving along different entropic states causes the emergence of a natural orientation in terms of information (an arrow of time). This behavior resembles the conceptual idea of the phenomenon known as hysteresis.
}
\label{fig:FisherCurve_beta}
\end{figure}

In practical terms, what happens to the metric tensor can be summarized as: by moving forward $\delta$ units in the $\beta$ ``axis'' we sense an effect that is not always the inverse of the effect produced by a displacement of $-\delta$ units in the opposite direction. In other words, moving towards higher entropy states (when $\beta$ increases) is different from moving towards lower entropy states (when $\beta$ decreases). This effect, which resembles the conceptual idea of a hysteresis phenomenon \cite{Hysteresis}, in which the future output of the system depends on its history, is illustrated by a plot of the Fisher curve of the random field along the simulation. Making a analogy with a concrete example, it is like the parametric space were made of a plastic material, that when pressured by a force deforms itself. However, when the pressure is vanishing, a different deformation process takes place to recover the original shape. Figs. \ref{fig:FisherCurve_beta} shows the estimated Fisher curves $\vec{F}_{A}^{B}(\beta) = \left(I_{\beta\beta}^{(1)}(\vec{\theta}), I_{\beta\beta}^{(2)}(\vec{\theta}), H_{\beta} \right)$ for $\beta=0,\ldots,0.5$ (the blue curve) and $\vec{F}_{B}^{A}(\beta) = \left(I_{\beta\beta}^{(1)}(\vec{\theta}), I_{\beta\beta}^{(2)}(\vec{\theta}), H_{\beta} \right)$ for $\beta=0.5,\ldots,0$ (the red curve) regarding each component of the metric tensor.

This natural orientation in the information space induces an arrow of time along the evolution of the random field. In other words, the only way to go from A to B by the red path would be running the simulation backwards. Note, however, that when moving along states whose variation in entropy is negligible (for example, a state A' in the same plane of constant entropy) the notion of time is not apparent. In other words, it is not possible to know whether we are moving forward or backwards in time, simply because at this point the notion of time is not clear (time behaves similar to a space-like dimension since it is possible to move in both directions in this information space, once the states A and A' are equivalent in terms of entropy, because there is no significant variation of $H_{\beta}$). During this period, it the perception of the passage of time is not clear, since the deformations induced by the metric
tensor into the parametric space (manifold) are reversible for opposite displacements in the inverse temperature direction. Note also that, from a differential geometry perspective, the torsion of the curve seems to be related to the divergence between the two types of Fisher information. When $I_{\beta\beta}^{(1)}(\vec{\theta})$ diverges from $I_{\beta\beta}^{(2)}(\vec{\theta})$ the Fisher curve leaves the plane of constant entropy. The results suggest that the torsion of the curve at a given point could be related to the notion of the passage of time: large values suggest that time seems to be "running faster" (large change in entropy) while small values suggest the opposite (if we are moving through a plane of constant entropy then time seems to be "frozen").     

\begin{figure}[h]
\begin{center}
\includegraphics[scale=1]{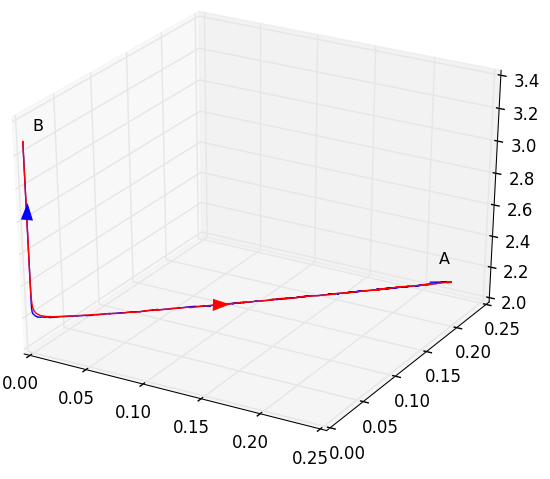}
\end{center}
\caption{{\bf Fisher curve of the random field regarding the component $I_{\mu\mu}(\vec{\theta})$.} 
The parametric curve was built by varying the inverse temperature parameter $\beta$ from $\beta_{MIN}=0$ (state A) to $\beta_{MAX}=0.5$ (state B) and back. In this case the arrow of time is not evident since the two curves, $\vec{F}_{A}^{B}(\beta)$ and $\vec{F}_{B}^{A}(\beta)$, are essentially the same.
}
\label{fig:FisherCurve_mu}
\end{figure}

\begin{figure}[h]
\begin{center}
\includegraphics[scale=1]{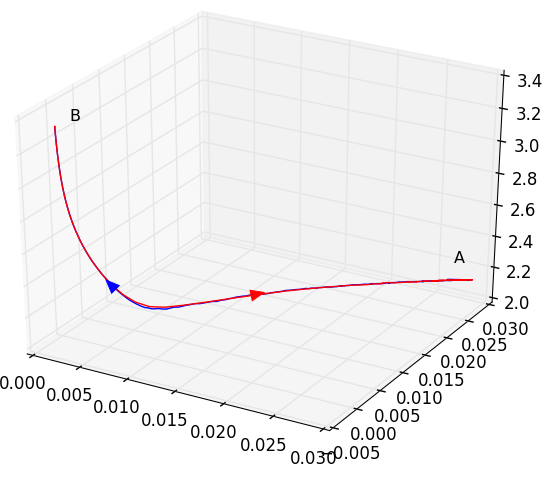}
\end{center}
\caption{{\bf Fisher curve of the random field regarding the component $I_{\sigma^{2}\sigma^{2}}(\vec{\theta})$.} 
The parametric curve was built by varying the inverse temperature parameter $\beta$ from $\beta_{MIN}=0$ (state A) to $\beta_{MAX}=0.5$ (state B) and back. In this case the arrow of time is not evident since the two curves, $\vec{F}_{A}^{B}(\beta)$ and $\vec{F}_{B}^{A}(\beta)$, are essentially the same.
}
\label{fig:FisherCurve_sigma}
\end{figure}

\begin{figure}[h]
\begin{center}
\includegraphics[scale=1]{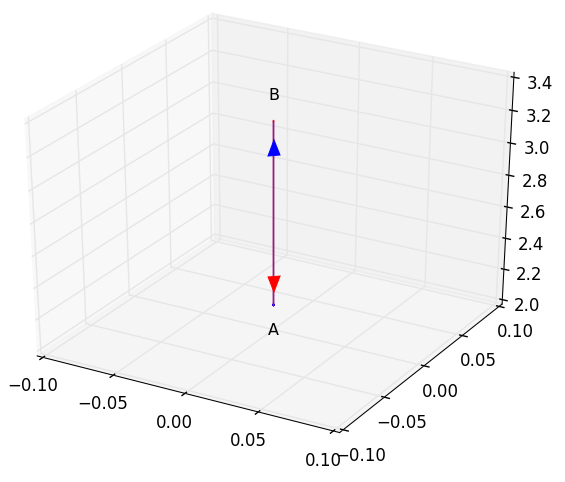}
\end{center}
\caption{{\bf Fisher curve of the random field regarding the component $I_{\sigma^{2}\beta}(\vec{\theta})$.}
The parametric curve was built by varying the inverse temperature parameter $\beta$ from $\beta_{MIN}=0$ (state A) to $\beta_{MAX}=0.5$ (state B) and back. Once again, in this case the arrow of time is not evident since the two curves, $\vec{F}_{A}^{B}(\beta)$ and $\vec{F}_{B}^{A}(\beta)$, are essentially the same.
}
\label{fig:FisherCurve_sigbeta}
\end{figure}

Following the same strategy, the Fisher curves regarding the remaining components were generated. Figs. \ref{fig:FisherCurve_mu}, \ref{fig:FisherCurve_sigma} and \ref{fig:FisherCurve_sigbeta} illustrates the obtained results. Note, however, that the notion of time is not captured in these curves. By looking at these measurements we cannot say whether the system is moving forwards or backwards in time, even for large variations on the inverse temperature parameter. Since the Fisher curves $\vec{F}_{A}^{B}(\beta)$ and $\vec{F}_{B}^{A}(\beta)$ are essentially the same, the path from A ($\beta = 0$) to B ($\beta=0.5$) is the inverse of the path from B to A.

\section{Discussion}
This section describes the main results obtained in this paper, focusing on the interpretation of the
proposed mathematical model of hysteresis for the study of complex systems: the Fisher curve of a random field. Basically, when temperature is infinite ($\beta = 0$) entropy fluctuates around a minimum base value and the information equality prevails. From an information geometry perspective,
a reduction in temperature (increase in $\beta$) causes a series of changes in the geometry of the parametric space, since the metric tensor (Fisher information matrix) is drastically deformed in an apparently non-reversible way, inducing the emergence of a natural orientation of evolution (arrow of time).

By quantifying and measuring an arrow of time in random fields, a relevant aspect that naturally arises concerns the notions of past and future. Suppose the random field is now in a state A, moving towards an increase in entropy (that is, $\beta$ is increasing). Within this context, the analysis of the Fisher curves suggests a possible interpretation: past is a notion related to a set of states $P = \left\{ X^{(\beta-)} \right\}$ whose entropies are below the current entropic plane of the state A. Equivalently, the notion of past could also be related to a set of states $P = \left\{ X^{(\beta+)} \right\}$ whose entropies are above the current entropic plane, provided the random field is moving towards a lower entropy state.

\begin{figure}[h]
\begin{center}
\includegraphics[scale=0.7]{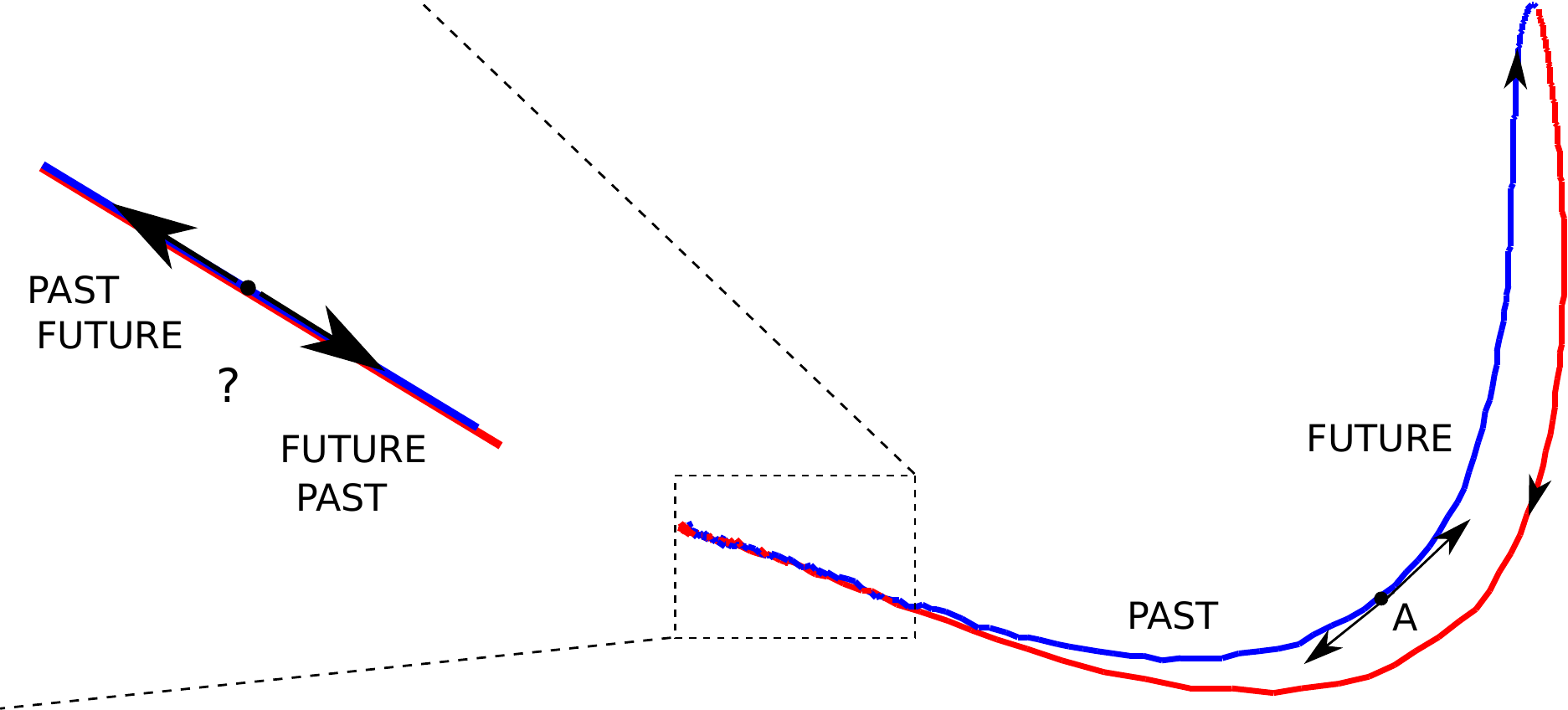}
\end{center}
\caption{{\bf The Fisher curve, an arrow of time and notions of past and future in the evolution of a Gaussian random field.} 
When there is no significant change in entropy, the deformations in the metric tensor components are reversible and therefore the arrow of time is not visible (no hysteresis is observed)
}
\label{fig:PAST_FUTURE}
\end{figure}

Again, let us suppose the random field is in a state A and moving towards an increase in entropy ($\beta$ is increasing). Similarly, the notion of future refers to a set of states $F = \left\{ X^{(\beta+)} \right\}$ whose entropies are higher than the entropy of the current state A (or equivalently, future could also refer to the set of states $F = \left\{ X^{(\beta-)} \right\}$ whose entropies are lower than A, provided that the random field is moving towards a decrease in entropy). According to this possible interpretation, the notion of future is related to the direction of the movement, pointed by the tangent vector at a given point of the Fisher curve. If along the evolution of the random field there is no significant change in the system's entropy, then time behaves similar to a spatial dimension, as illustrated by Fig. \ref{fig:PAST_FUTURE}.

\section{Conclusions}
In this paper, we addressed the problem of characterizing the emergence of an arrow of time in Gaussian random field models. To intrinsically investigate the effect of the passage of time, we performed computational simulations in which the inverse temperature parameter is controlled to guide the system behavior throughout different entropic states. Investigations about the relation between two important information theoretic measures, entropy and Fisher information, led us to the definition of the Fisher curve of a random field, a parametric trajectory embbeded in an information space, which characterizes the system behavior in terms of variations in the metric tensor of the statistical manifold. Basically, this curve provides a geometrical tool for the analysis of random fields by showing how different entropic states are "linked" in terms of Fisher information, which is, by definition, the metric tensor of the underlying random field model parametric space. In other words, when the random field moves along different entropic states, its parametric space is actually being deformed by changes that happen in Fisher information matrix (the metric tensor). In this scientific investigation we observe what happens to this geometric structure when the inverse temperature parameter is modified, that is, when temperature deviates from infinity, by measuring both entropy and Fisher information. An indirect subproblem involved in the solution of this main problem was the estimation of the inverse temperature parameter of a random field, given an outcome (snapshot) of the system. To tackle this subproblem, we used a statistical approach known as maximum pseudo-likelihood estimation, which is especially suitable for random fields, since it avoids computations with the joint Gibbs distribution, often computationally intractable. Our obtained results show that moving towards higher entropy states is different from moving towards lower entropy states, since the Fisher curves are not the same. This asymmetry induces a natural orientation to the process of taking the random field from an initial state A to a final state B and back, which is basically the direction pointed by the arrow of time, since the only way to move in the opposite direction is by running the simulations backwards. In this context, the Fisher curve can be considered a mathematical model of hysteresis in which the natural orientation is given by the arrow of time. Future works may include the study of the Fisher curve in other random field models, such as the Ising and q-state Potts models.


\begin{thebibliography}{64}

\bibitem{CPLX}
Chu D, Strand R, Fjelland R., Theories of complexity. Complexity. 8(3):19--30, 2003.

\bibitem{Sibani}
Sibani P, Jensen HJ., Stochastic Dynamics of Complex Systems. World Scientific. 2013.

\bibitem{Strogatz}
Strogatz SH. Exploring complex networks. Nature. 2001;410:268--276.

\bibitem{Barabasi}
Albert R, Barab\'asi AL. Statistical mechanics of complex networks. Rev Mod Phys. 2002 Jan;74:47--97.

\bibitem{Newman2003}
Newman MEJ. The Structure and Function of Complex Networks. SIAM Review. 2003;45(2):167--256.

\bibitem{Boccaletti}
Boccaletti S, Latora V, Moreno Y, Chavez M, Hwang DU. Complex networks: Structure and dynamics. Physics Reports. 2006;424(4–5):175 -- 308.

\bibitem{BigData}
Hassanien AE, Taher~Azar A, Snasel V, Kacprzyk J, Abawajy JHE.
\newblock Big Data in Complex Systems: Challenges and Opportunities.
\newblock Springer; 2015.

\bibitem{Hastie}
Hastie T, Tibshirani R, Friedman J.
\newblock The Elements of Statistical Learning: Data Mining, Inference, and
  Prediction.
\newblock 2nd ed. Springer; 2009.

\bibitem{Mining}
Wu X, Zhu X, Wu GQ, Ding W.
\newblock Data mining with big data.
\newblock Knowledge and Data Engineering, IEEE Transactions on. 2014
  Jan;26(1):97--107.

\bibitem{escience}
Jankowski NW.
\newblock Exploring e-Science: An Introduction.
\newblock Journal of Computer-Mediated Communication. 2007;12(2):549--562.
\newblock Available from:

\bibitem{RandomFields}
Vanmarcke E.
\newblock Random Fields: Analysis and Synthesis.
\newblock World Scientific; 2010.

\bibitem{Besag1974}
Besag J.
\newblock Spatial interaction and the statistical analysis of lattice systems.
\newblock Journal of the Royal Statistical Society - Series B.
  1974;36:192--236.

\bibitem{MLaPP}
Murphy KP.
\newblock Machine Learning: a Probabilistic Perspective.
\newblock MIT Press; 2012.

\bibitem{Boltzmann}
Boltzmann L.
\newblock On certain questions of the Theory of Gases.
\newblock Nature. 1895;51:413--415.

\bibitem{Gibbs}
Gibbs JW.
\newblock Elementary principles in statistical mechanics.
\newblock Charles Scribner's Sons; 1902.

\bibitem{Ising1925}
Ising E.
\newblock Beitrag zur Theorie des Ferromagnetismus.
\newblock Zeitschr f Physik. 1925;39:253--258.

\bibitem{Heisenberg}
Heisenberg W.
\newblock Zur Theorie des Ferromagnetismus.
\newblock Zeitschr f Physik. 1928;49(9-10):619--636.

\bibitem{Spitzer}
Spitzer F.
\newblock Markov random fields and Gibbs ensembles.
\newblock The American Mathematical Monthly. 1971;78:142--154.

\bibitem{PottsWu}
Wu FY.
\newblock The Potts Model.
\newblock Reviews of Modern Physics. 1982;54(1):235--268.

\bibitem{Albeverio}
Albeverio S, Hoegh-Krohn R.
\newblock Homogeneous random fields and statistical mechanics.
\newblock Journal of Functional Analysis. 1975;19(3):242--272.

\bibitem{Montroll1941}
Montroll E.
\newblock Statistical mechanics of nearest neighbor systems.
\newblock Journal of Chemical Physics. 1941;9:706.

\bibitem{Onsager}
Onsager L.
\newblock Crystal statistics. 1. A two-dimensional model with an order-disorder
  transition.
\newblock Physical Review. 1944;65(3-4):117--149.

\bibitem{KnotWu}
Wu FY.
\newblock Jones polynomial as a Potts model partition function.
\newblock Journal knot Theory and Ramifications. 1992;1(1):47--57.

\bibitem{Liu}
Liu Z, Luo J, Shao C.
\newblock Potts model for exaggeration of a simple rumor transmitted by
  recreant rumormongers.
\newblock Physical Review E. 2001;64(6):046134 [9 pages].

\bibitem{Merks}
Merks RMH, Glazier JA.
\newblock A cell-centered approach to developmental biology.
\newblock Physica A: Statistical Mechanics and Applications.
  2005;352(1):113--130.

\bibitem{Geman1984}
Geman S, Geman D.
\newblock Stochastic relaxation, {Gibbs} distributions, and the {Bayesian}
  restoration of images.
\newblock IEEE Trans on Pattern Analysis Machine Intelligence.
  1984;6(6):721--741.

\bibitem{Sabelfeld}
Sabelfeld K. K.
\newblock Random Fields and Stochastic Lagrangian Models. Analysis and Applications in Turbulence and Porous Media.
\newblock Walter de Gruyter, Berlin, 2012, 399 pp. ISSN 978-3-11-029681-5

\bibitem{Amari1985}
Amari S.
\newblock Differential-geometrical methods in statistics (Lecture notes in
  statistics).
\newblock Springer-Verlag; 1985.

\bibitem{Amari2000}
Amari NH S.
\newblock Methods of information geometry (Translations of mathematical
  monographs v. 191).
\newblock American Mathematical Society; 2000.

\bibitem{Frieden2004}
Frieden BR.
\newblock Science from Fisher Information: A Unification.
\newblock Cambridge: Cambridge University Press; 2004.

\bibitem{Dodson}
Arwini KA, Dodson CTJ.
\newblock Information Geometry: Near Randomness and Near Independence.
\newblock Springer; 2008.

\bibitem{Fisher}
Fisher RA.
\newblock Theory of Statistical Estimation.
\newblock Mathematical Proceedings of the Cambridge Philosophical Society. 1925;22(5):700--725
\newblock Available from:

\bibitem{Nielsen}
Barndorff-Nielsen OE.
\newblock Information and Exponential Families in Statistical Theory.
\newblock Wiley; 1978.

\bibitem{Pistone}
Pistone G, Rogantin MP.
\newblock The exponential statistical manifold: Mean parameters, orthogonality,
  and space transformation.
\newblock Bernoulli. 1999;5:721--760.

\bibitem{Amari2001}
Amari SI.
\newblock Information Geometry on Hierarchy of Probability Distributions.
\newblock IEEE Transactions on Information Theory. 2001;47(5):1701--1711.

\bibitem{Amari1992}
Amari SI, Kurata K, Nagaoka H.
\newblock Information geometry of Boltzmann machines.
\newblock Neural Networks, IEEE Transactions on. 1992 Mar;3(2):260--271.

\bibitem{Janke}
Janke W, Johnston DA, Kenna R.
\newblock Information Geometry and Phase Transitions.
\newblock Physica A. 2004;336:181--186.

\bibitem{Xavier}
Calmet X, Calmet J.
\newblock Dynamics of the Fisher Information Metric.
\newblock Physical Review E. 2005;71:056109.

\bibitem{Zanardi}
Zanardi P, Giorda P, Cozzini M.
\newblock Information-Theoretic Differential Geometry of Quantum Phase
  Transitions.
\newblock Physical Review Letters. 2007;99:100603.

\bibitem{Campisi}
Campisi M, Hanggi P.
\newblock Fluctuation, Dissipation and the Arrow of Time.
\newblock Entropy. 2011;13(12):2024--2035.

\bibitem{TimeArrow}
Jejjala V, Kavic M, Minic D, Tze CH.
\newblock Modeling Time's Arrow.
\newblock Entropy. 2012;14(4):614--629.

\bibitem{Haddad}
Haddad WM.
\newblock Temporal Asymmetry, Entropic Irreversibility, and Finite-Time
  Thermodynamics: From Parmenides-Einstein Time-Reversal Symmetry to the
  Heraclitan Entropic Arrow of Time.
\newblock Entropy. 2012;14(3):407--455.

\bibitem{Sueli}
Costa SIR, Santos SA, Strapasson JE.
\newblock Fisher information distance: A geometrical reading.
\newblock Discrete Applied Mathematics. 2014

\bibitem{Odemir}
Machicao, J., Marco A. G., Bruno, O. M.
\newblock Chaotic encryption method based on life-like cellular automata
\newblock Expert Systems with Applications. 2012;39(16):12626--12635 

\bibitem{Levada2014}
Levada ALM.
\newblock Learning from Complex Systems: On the Roles of Entropy and Fisher
  Information in Pairwise Isotropic Gaussian Markov Random Fields.
\newblock Entropy, Special Issue on Information Geometry. 2014;16:1002--1036.

\bibitem{Hastings1970}
Hastings WK.
\newblock Monte Carlo sampling methods using Markov chains and their
  applications.
\newblock Biometrika. 1970;57:97--109.

\bibitem{Swendsen1987}
Swendsen R, Wang J.
\newblock Nonuniversal critical dynamics in Monte Carlo simulations.
\newblock Physical Review Letters. 1987;58:86--88.

\bibitem{Wolff1989}
Wolff U.
\newblock Collective Monte Carlo updating for spin systems.
\newblock Physical Review Letters. 1989;62:361--364.

\bibitem{gilks1993}
Gilks WR, Clayton DG, Spiegelhalter DJ, Best NG, McNeil AJ, Sharples LD, et~al.
\newblock Modeling complexity: Applications of Gibbs sampling in medicine.
\newblock {Journal of the Royal Statistical Society, Series B}.
  1993;55(1):39--52.

\bibitem{smith1993}
Smith AFM, Robert GO.
\newblock Bayesian computation via the Gibbs sampler and related Markov chain
  Monte Carlo methods.
\newblock {Journal of the Royal Statistical Society, Series B}.
  1993;55(1):3--23.

\bibitem{Roberts1996}
Roberts GO.
\newblock Markov chain concepts related to sampling algorithms.
\newblock In: Gilks WR, Richardson S, Spiegelhalter DJ, editors. Markov Chain
  Monte Carlo in practice \normalfont \textsc{(edited by Gilks, W. R.,
  Richardson, S. and Spiegelhalter, D. J.)}. Chapman \& Hall/CRC; 1996. p.
  45--57.

\bibitem{Landau2000}
Landau DP, Binder K.
\newblock A guide to monte carlo simulations in statistical physics.
\newblock Cambridge: Cambridge University Press; 2000.

\bibitem{Chib2004}
Chib S.
\newblock Makov Chain Monte Carlo Technology.
\newblock In: J~E~Gentle WH, Mori Y, editors. Handbook of Computational
  Statistics \normalfont \textsc{(edited by J. E. Gentle, W. H\"ardle and Y.
  Mori)}. Springer; 2004. p. 72--98.

\bibitem{Hammersley1971}
Hammersley JM, Clifford P.
\newblock {Markov} field on finite graphs and lattices; 1971.
\newblock Unpublished.

\bibitem{Kass1989}
Kass RE.
\newblock The Geometry of Asymptotic Inference.
\newblock Statistical Science. 1989;4(3):188--234.

\bibitem{isserlis1918}
Isserlis L.
\newblock On a formula for the product-moment coefficient of any order of a
  normal frequency distribution in any number of variables.
\newblock Biometrika. 1918;12:134--139.

\bibitem{Silvey}
Silvey SD.
\newblock Statistical Inference.
\newblock Chapman \& Hall/CRC Monographs on Statistics \& Applied Probability;
  1975.

\bibitem{Casella2002}
Casella G, Berger RL.
\newblock Statistical Inference.
\newblock 2nd ed. New York: Duxbury; 2002.

\bibitem{shannon1949}
Shannon C, Weaver W.
\newblock The Mathematical Theory of Communication.
\newblock University of Illinois Press, Urbana, Chicago, IL \& London; 1949.

\bibitem{jaynes1957}
Jaynes E.
\newblock Information theory and statistical mechanics.
\newblock Physical Review. 1957;106:620--630.

\bibitem{Jensen}
Jensen JL, K\"unsh HR.
\newblock On asymptotic normality of pseudo likelihood estimates for pairwise
  interaction processes.
\newblock Annals of the Institute of Statistical Mathematics.
  1994;46(3):475--486.

\bibitem{winkler}
Winkler G.
\newblock Image Analysis, Random Fields and Markov Chain Monte Carlo Methods: A
  Mathematical Introduction.
\newblock Secaucus, NJ, USA: Springer-Verlag New York, Inc.; 2006.

\bibitem{Covariance}
Liang G, Yu B.
\newblock Maximum pseudo likelihood estimation in network tomography.
\newblock IEEE Trans on Signal Processing. 2003;51(8):2043--2053.

\bibitem{Metropolis1953}
Metropolis N, Rosenbluth AW, Rosenbluth MN, Teller AH, Teller E.
\newblock Equation of state calculations by fast computing machines.
\newblock Journal of Chemical Physics. 1953;21(6):1087--1092.

\bibitem{Hysteresis}
Mayergoyz ID.
\newblock Mathematical Models of Hysteresis and their Applications.
\newblock Academic Press; 2003.

\end{thebibliography}
\end{document}